\documentclass[aps,twocolumn,showpacs,preprintnumbers,letter]{revtex4}
\usepackage{graphicx}

\newcommand{\beq}{\begin{equation}}
\newcommand{\eeq}{\end{equation}}
\newcommand{\be}{\begin{equation}}
\newcommand{\ee}{\end{equation}}
\newcommand{\ber}{\begin{eqnarray}}
\newcommand{\eer}{\end{eqnarray}}
\newcommand{\berr}{\begin{eqnarray*}}
\newcommand{\eerr}{\end{eqnarray*}}
\newcommand{\ba}{\begin{array}}
\newcommand{\ea}{\end{array}}

\newcommand{\sg}{{g_l(r,r',E+i\epsilon)}} 

\newcommand{\old}[1]{}

\begin{document}
\title{Can quarkonia survive deconfinement?}

\author{\'Agnes M\'ocsy \email{mocsy@bnl.gov}} \affiliation{ RIKEN-BNL
  Research Center, Brookhaven National Laboratory, Upton NY 11973, USA }
\author{P\'eter Petreczky\email{petreczk@bnl.gov}} \affiliation{
  RIKEN-BNL Research Center and Physics Department, Brookhaven
  National Laboratory, Upton NY 11973, USA }

\begin{abstract}
We study quarkonium correlators and spectral functions at zero and
finite temperature in QCD with only heavy quarks using potential
models combined with perturbative QCD. First, we show that this approach can
describe the quarkonium correlation function at zero temperature. 
Using a class of screened potentials based on lattice calculations of the static 
quark-antiquark free energy we calculate spectral functions at finite temperature.
We find that all quarkonium states, with the exception of the $1S$ bottomonium,
dissolve in the deconfined phase at temperatures smaller than $1.5T_c$,  in contradiction
with the conclusions of recent studies. Despite this the temperature dependence of
the quarkonium correlation functions calculated on the lattice is well reproduced in our model. 
We also find that even in the absence of
resonances the spectral function at high temperatures is significantly
enhanced over the spectral function corresponding to free quark
antiquark propagation.
\end{abstract}
\preprint{BNL-NT  07/21}
\preprint{RBRC-677 }
\maketitle

\section{introduction}
\label{sec:intro}

The study of quarkonia at finite temperature is interesting for
several reasons. First, due to their small size, these heavy
quark-antiquark bound states provide a bridge between perturbative
and nonperturbative Quantum Chromodynamics (QCD), testing forces at
intermediate distances.  At high temperatures, color screening, which
is usually understood in terms of in-medium modification of
interquark forces, occurs.  Based on this Matsui and Satz argued that
above the transition temperature $T_c$ screening effects are strong
enough to lead to dissolution of the $J/\psi$ state. This can then be
used as a signal of quark gluon plasma formation in heavy ion
collisions \cite{MS86}.

Because of the large quark mass $m = m_{c,b} \gg \Lambda_{QCD}$, the
velocity $v$ of heavy quarks in the bound state is small, and the
binding effects in quarkonia at zero temperature can be understood in
terms of a nonrelativistic potential model with a Coulomb plus linear
form, known as the Cornell potential
\cite{Eichten75,Eichten78,Eichten80,Buchmuller80}.  Potential models
appeared to be very successful in describing the quarkonium spectrum
and have been extensively used in the past 20 years, see
Ref. \cite{Lucha91}.  More recently, an understanding has developed on
how to derive the potential models from QCD using a sequence of
effective field theories : nonrelativistic QCD (NRQCD), an effective
theory where all modes above the scale $m$ are integrated out, and
potential nonrelativistic QCD (pNRQCD), an effective theory in which
all modes above the scale $mv $ are integrated out
\cite{Pineda98,Brambilla00}.  The concept of the potential can be
given a solid field theoretical definition in this framework at any
order of perturbation theory. In particular, the quark antiquark
potential is defined in terms of the expectation value of the Wilson
loops. Relativistic corrections to the potential can also be
calculated. They are expressed in terms of the Wilson loops with
appropriate insertions of electric and magnetic fields, see
Refs. \cite{Bali01,Brambilla05,yellow}.

Based on the success of the potential model at zero temperature, and
the idea that color screening implies modification of the interquark
forces, attempts to understand quarkonium properties at finite
temperature using potential models have been made
\cite{MS86,karsch88,hashimoto88,ropke88}. While in these works
phenomenological potentials have been used, more recent studies went
one step further and attempted to connect the potential to lattice
calculations of the finite temperature free energy of a static quark
antiquark pair
\cite{digal01a,digal01b,shuryak04,wong04,wong06,alberico,rapp,alberico06}.

Quarkonium states at zero temperature are well defined: their widths are
very small compared to their binding energies, at least for states below
the threshold. At finite temperature the situation is different. We
expect that all quarkonium states will acquire a sizable width, which
increases with increasing temperature.  At some temperature the width
becomes large enough that it is no longer meaningful to talk about
individual quarkonium states. Instead, one should consider the
spectral function, which contains contributions from all possible
states in a channel with given quantum numbers.  Furthermore, the spectral function
in the vector channel is a quantity which can be measured directly, since it
is proportional to the dilepton production rate. Quarkonium spectral
functions at finite temperature have been considered only relatively
recently. Using lattice QCD, charmonium correlators have been
calculated and the corresponding spectral functions have been
extracted using the Maximum Entropy Method (MEM)
\cite{umeda02,asakawa04,datta04,datta_sewm04,datta_panic05,doi,swan,jako06}.
Although this approach can in principle provide the ultimate solution
to the problem of in-medium quarkonium properties, current
calculations have serious limitations and cannot give detailed
information about quarkonium spectral functions.  Using the MEM at
zero temperature one can reconstruct the basic features of the
spectral functions: the ground state, the excited states, and the
continuum \cite{jako06}.  Individual excited states, however, cannot
be resolved. At finite temperature, even resolving the ground state
appears to be difficult with existing lattice data.  The only
statement that can be made at this time, is that in the pseudoscalar
channel, quarkonium spectral functions do not show significant changes
up to temperatures as high as $1.5T_c$, while in the scalar channel
the spectral function is strongly modified just above the transition
temperature \cite{jako06}.

For this reason, quarkonium correlation functions have been studied
using a simplified model of the spectral function, which contained
bound states and a perturbative continuum \cite{mocsy05,mocsy06}.  The
results of these calculations have been compared to lattice results
and no agreement has been found. Very recently this approach has been
extended using the full nonrelativistic Green's function of heavy quark
antiquark pairs to estimate the spectral function and the
corresponding Euclidean correlators
\cite{rapp,agi_proc,alberico06}.  However, even in these approaches no agreement with lattice
calculations has been found.

In the present paper we study spectral functions in the pseudoscalar,
vector, scalar and axial-vector channels which correspond to
$\eta_{c}$ ($\eta_b$), $J/\psi$ ($\Upsilon$),  $\chi_{c0}$
($\chi_{b0}$) and $\chi_{c1}$ ($\chi_{b1}$) charmonium (bottomonium)
states, respectively.  In the energy region below and near the
continuum threshold the spectral function is calculated using a
potential model and nonrelativistic Green's function. Since the potential at finite temperature is not known, 
we consider a class of screened potentials based on lattice results on the static quark-antiquark free energy.  
Well above the threshold, the nonrelativistic spectral function is matched to the
perturbative fully relativistic result. From this the Euclidean
correlators are calculated. We then compare these correlators to the
results of recent lattice calculations. We find that the lattice data
does not necessarily imply survival of different quarkonium
states. Rather, despite the fact that most quarkonia, with the
exception of the 1S bottomonium, are dissolved at temperatures smaller
than $1.5T_c$, agreement with the correlator data is found. Clearly,
dissociation temperatures previously quoted in the literature (e.g. $2T_c$
for $J/\psi$ ) have not  been seen in our analysis.

The rest of the paper is organized as follows: The framework for
calculating quarkonium spectral functions is discussed in Section
II. In Section III we show our analysis of the quarkonium spectral
functions and Euclidean correlators at zero temperature and compare to
lattice QCD results. Sections IV and V contain our results on the
finite temperature spectral function and correlators. Finally, in
Section VI we give our conclusions and outlook. The reader not
interested in technical details can skip Sections II and III. Further
technical details of our calculations are presented in the Appendices.

\section{Quarkonium spectral functions in the nonrelativistic limit}
\label{sec:nrspf}

In this section we discuss quarkonium spectral functions in the free theory as well
as in the  nonrelativistic limit.
In what follows we consider the case of zero spatial momentum, i.e. quarkonium at rest.  
The spectral function is defined as the
imaginary part of the retarded current-current correlator
\begin{eqnarray}
&
\displaystyle
\sigma(\omega)=-\frac{1}{\pi} {\rm Im} D_R(\omega), \\
&
\displaystyle
i D_R(\omega)=\int dt e^{i \omega t} \theta(t) \int d^3 x  \langle [j(t,\vec{x}), j(0,0)] \rangle,  
\end{eqnarray}
It carries information about all the possible states with a given
quantum number, which is fixed by the current
\begin{equation}
j=\bar q \Omega q
\end{equation}
with $\Omega=1,\gamma_5, \gamma_{\mu}, \gamma_{\mu} \gamma_5$ for scalar, pseudoscalar,
vector and axial vector channels respectively.

For large $\omega$ the spectral function can be
calculated in perturbation theory (see e.g. \cite{karsch03,aarts05})
\begin{equation}
\sigma(\omega)=\frac{N_c}{8 \pi^2} \omega^2 \left( a + b \frac{s_0^2}{\omega^2} \right )
\sqrt{1-\frac{s_0^2}{\omega^2}} \left ( 1 + C \frac{\alpha_s}{\pi} \right ).
\label{free_rel_spf}
\end{equation}
Here $a=1,~b=0$ for pseudoscalar, $a=2,~b=3$ for vector, $a=1,b=-1$
for the scalar and $a=2,b=-2$ in the axial-vector channel,
respectively. In leading order perturbation theory the threshold is
$s_0=2 m_{c,b}$. The coefficient $C$ of the leading perturbative
correction has been calculated only for the massless case \cite{reinders84}. 
The number of colors in QCD is
$N_c=3$.  

While perturbation theory is reliable away from the
threshold, $\omega \gg s_0$, the physics becomes quite complicated
near the threshold, even in the weak coupling regime
\cite{braun,khoze,peskin91,laine04}. Close to the threshold the quark
and antiquark move slowly allowing enough time for multiple gluon
exchange. Adding an extra gluon exchange does not lead to a
suppression by $\alpha_s$. In this case we need to resum ladder
diagrams. In the following we discuss this resummation separately for
the pseudoscalar and scalar channels. The vector channel has been
discussed in detail in Ref. \cite{peskin91}, while the axial-vector
channel is completely analogous to the scalar case.

\subsection{pseudoscalar channel}

The summation of ladder diagrams corresponds to solving the integral
equation for the vertex function
\begin{equation}
\Gamma(p,q)=\gamma_5+\int \frac{d^4 k}{(2 \pi)^4} S_F(k+\frac{q}{2}) \Gamma(k,q) S_F(k-\frac{q}{2}) \tilde V(p-k)
\end{equation} 
and inserting the solution of this equation into the 1-loop expression
of the meson correlator (see e.g. \cite{peskin91})
\begin{equation}
D(q^2)=N_c \int \frac{d^4 p}{(2 \pi)^4} Tr \left[ S_F(p+\frac{q}{2})
  \Gamma(p,q) S_F(p-\frac{q}{2}) \gamma_5 \right ].
\label{Dq2}
\end{equation}
In general this task is complicated, but it simplifies considerably in the
nonrelativistic limit.  In this case following Ref. \cite{peskin91}
we can replace the quark propagators with the corresponding
nonrelativistic forms
\begin{eqnarray}
&
\displaystyle
S_F\left(p+\frac{q}{2}\right)=\frac{(1+\gamma_0)m+\gamma_0 (p_0+E/2)+{\vec{\gamma}}\cdot {\vec{p}}}
{2 m(\frac{E}{2}+p_0-\frac{\vec{p}^2}{2m}+i \epsilon/2)},\nonumber\\[3mm]
&
\displaystyle
S_F\left(p-\frac{q}{2}\right)=\frac{(1-\gamma_0)m+\gamma_0 (p_0-E/2)+\vec{\gamma}\cdot \vec{p}}
{2 m(\frac{E}{2}-p_0-\frac{\vec{p}^2}{2m}+ i \epsilon/2)}.
\label{quark_prop_nr}
\end{eqnarray}
Here we have taken into account that $q=(2m+E,\vec{0})$ and $E \ll 2
m$. The 1-gluon exchange operator in this limit is $\tilde
V(\vec{k}-\vec{p}) \equiv V(|\vec{k}-\vec{p}|)=\frac{4}{3} \alpha_s 4
\pi D_{00}(|\vec{k}-\vec{p}|)$, where $D_{00}$ is the temporal Coulomb
gauge gluon propagator.  In this limit the vertex function can be
chosen to be independent of $p_0$ \cite{peskin91}.

The leading order result for the pseudoscalar spectral function can
be obtained by replacing $\Gamma=\gamma_5$ in Eq. (\ref{Dq2}) and
using the nonrelativistic form of the quark propagators above with
only the first term in the denominator.  The retarded nature of the
meson correlator $D(q^2)$ is ensured by the $+i \epsilon$ prescription
in the quark propagators.  This gives for the free nonrelativistic
spectral function
\begin{equation}
\sigma(E)=-\frac{1}{\pi} {\rm Im} D(q^2)=\frac{N_c}{2 \pi^2} m^{3/2} E^{1/2}.
\end{equation}
Note, that this result can also be obtained from
Eq. ({\ref{free_rel_spf}) by writing $\omega=2 m+E$ and $s_0=2 m$, and
  expanding in $E/m$ to leading order.

Defining the scalar function
\begin{equation}
\frac{(1+\gamma_0)}{2} \Gamma(\vec{p},E) \frac{(1-\gamma_0)}{2}=\frac{(1+\gamma_0)}{2} \gamma_5 \frac{(1-\gamma_0)}{2} 
\tilde \Gamma(\vec{p},E)
\end{equation}
and performing the integral over $k_0$ explicitly, the integral
equation for the vertex function can be written as
\begin{equation}
\tilde \Gamma(\vec{p},E)=1+ \int \frac{d ^3 k}{(2 \pi)^3} \frac{1}{E+i \epsilon -k^2/m} V(|\vec{p}-\vec{k}|) 
\tilde \Gamma(\vec{k}, E).
\end{equation}
By introducing the nonrelativistic Green's function 
\begin{equation}
\displaystyle
G^{nr}(\vec{k},E+ i \epsilon)=-\frac{1}{E+ i \epsilon-\frac{k^2}{m}} \tilde \Gamma(\vec{k},E),
\end{equation}
the equation for the scalar vertex function $\tilde \Gamma(\vec{k},E)$
can be rewritten in the form
\begin{eqnarray}
&
\displaystyle
-\left ( E + i \epsilon -\frac{p^2}{m} \right ) G^{nr}(\vec{p},E+i \epsilon)=\nonumber\\
&
\displaystyle
1-\int \frac{d^3 k}{(2 \pi)^3} V(|\vec{p}-\vec{k}|) G^{nr}(\vec{k},E+i \epsilon),
\label{schroedinger_mom}
\end{eqnarray}
which is the Schr\"odinger equation in momentum space. In a more
familiar form, in coordinate space, it reads
\begin{equation}
\left [ -\frac{1}{m} \vec{\nabla}^2+V(r)-(E+i \epsilon) \right ] G^{nr}(\vec{r},\vec{r'},E+i \epsilon) = \delta^3(r-r').
\label{schroedinger}
\end{equation}
Here 
\begin{equation}
G^{nr}(\vec{r},\vec{r'},E+i \epsilon)=\int \frac{d^3 k}{(2 \pi)^3} e^{-i \vec{k}\cdot (\vec{r}-\vec{r'})} 
G^{nr}(\vec{k},E+i \epsilon)
\end{equation}
is the nonrelativistic Green's function in coordinate space and 
\begin{equation}
V(r)=\int \frac{d^3 k}{(2 \pi)^3} V(|\vec{k}|) e^{-i \vec{k} \cdot \vec{r}}
\end{equation}
is the potential. Thus we can write
\begin{eqnarray}
&
\displaystyle
\sigma(E) = \frac{2 N_c}{\pi}{\rm Im} \int \frac{d^3 k}{(2 \pi)^3} G^{nr}(\vec{k},E+i \epsilon)=\nonumber\\
&
\displaystyle
\frac{2 N_c}{\pi} {\rm Im} G^{nr}(\vec{r},\vec{r'},E+i \epsilon)|_{\vec{r}=\vec{r'}=0}.
\label{green_ps}
\end{eqnarray}
Therefore, in order to calculate the pseudoscalar spectral function in
the nonrelativistic limit we have to solve the Schr\"odinger equation (\ref{schroedinger})
for $G^{nr}(\vec{r},\vec{r'},E+i \epsilon)$ and take the limit
$\vec{r}=\vec{r'}=0$, according to (\ref{green_ps}).

\subsection{Scalar Channel}
The calculation of the scalar spectral function is somewhat more
complicated. To understand the problem better, let us consider the
noninteracting case first. From the structure of the quark
propagators in the nonrelativistic limit it is clear that for the
scalar vertex $\Omega=1$ the meson correlator is equal to zero in
leading order of a $1/m$ expansion
(c.f. Eqs. (\ref{quark_prop_nr})). Therefore the second and third
terms in the numerator of Eqs. (\ref{quark_prop_nr}) should be
retained. We then get
\begin{eqnarray}
&
\displaystyle
D(q^2)=\nonumber\\
&
\displaystyle
N_c \int \frac{d^4 p}{(2 \pi)^4} \frac{\frac{\vec{p}^2}{m^2}-E/m}{(\frac{E}{2}+p_0+i \frac{\epsilon}{2}-\frac{\vec{p}^2}{2m})(\frac{E}{2}+p_0+i \frac{\epsilon}{2}-\frac{\vec{p}^2}{2m})}.
\nonumber\\
\end{eqnarray}
Taking the imaginary part and performing the integral we get
\begin{equation}
\sigma(E)=\frac{N_c}{2 \pi^2} m^{1/2} E^{3/2}
\end{equation}
Note that we arrive at the same result if we consider the
nonrelativistic scalar vertex $\displaystyle \Omega=\frac{\vec{\gamma} \cdot \vec{p}}{m}$
instead if the relativistic one $\Omega=1$. In fact, this type of
nonrelativistic vertex is used to study the $\chi_{c,b}$ states in
NRQCD \cite{lepage91}
\footnote{The relativistic quark field $\Psi$ can be written in terms
  of nonrelativistic quark and antiquark fields as
  $\Psi=(\psi,\chi)$. The simplest operator corresponding to scalar
  meson then is $\chi^{\dagger} \vec{\sigma}\vec{D} \psi/m$
  \cite{lepage91}. }.

Repeating all the steps discussed
in the previous section we write the correlator
\begin{equation}
D(q^2)=N_c \int \frac{d^4 p}{(2 \pi)^4} Tr \left[ S_F(p+\frac{q}{2})
  \Gamma(p,q) S_F(p-\frac{q}{2}) \frac{\vec{\gamma} \cdot \vec{p}}{
    m} \right ],
\label{Dq2sc}
\end{equation}
where the vertex function $\Gamma(p,q)$ satisfies the equation
\begin{equation}
\Gamma(p,q)=\frac{\vec{\gamma} \cdot \vec{p}}{m}+\int \frac{d^4 k}{(2 \pi)^4} S_F(k+\frac{q}{2}) 
\Gamma(k,q) S_F(k-\frac{q}{2}) V(p-k).
\end{equation} 
Introducing the scalar function 
\begin{equation}
\frac{(1+\gamma_0)}{2} \Gamma(p,q) \frac{(1-\gamma_0)}{2}=\frac{(1+\gamma_0)}{2}  \frac{\vec{\gamma} \cdot \vec{p}}{m} 
\frac{(1-\gamma_0)}{2} \tilde \Gamma(\vec{p},E)\, ,
\end{equation}
we write
\begin{equation}
D(q^2)=\int \frac{d ^3 k}{(2 \pi)^3} \frac{1}{E+i \epsilon -k^2/m} \tilde \Gamma(\vec{k}, E) \frac{\vec{k}^2}{m^2}\, .
\end{equation}
It is easy to see that $\tilde \Gamma(\vec{k}, E)$ satisfies
Eq. (\ref{schroedinger_mom}) and therefore we can write
\begin{eqnarray}
&
\displaystyle
\sigma(E)=-\frac{1}{\pi}{\rm Im} D(q^2)=\nonumber\\
&
\displaystyle
\frac{2 N_c}{\pi}\frac{1}{m^2} {\rm Im} \vec{\nabla} \cdot \vec{\nabla'} G^{nr}(\vec{r},\vec{r'},E+i \epsilon)|_{\vec{r}=\vec{r'}=0}.
\label{green_sc}
\end{eqnarray}
Thus to calculate the spectral function in the scalar channel, we have
to calculate the derivatives of the nonrelativistic Green's function
with respect of $\vec{r}$ and $\vec{r'}$ and take then the limit
$\vec{r}=\vec{r'}=0$, according to (\ref{green_sc}).

\subsection{Numerical Solution of the Schr\"odinger Equation}

To obtain the spectral function in the nonrelativistic limit, we have
to solve Eq. (\ref{schroedinger}) for nonzero $\epsilon$,
i.e. complex energy.  We use the numerical method developed in
Ref. \cite{peskin91} for this purpose which we have extended to
incorporate the scalar channel, as discussed in Appendix A.  At finite
temperature all particles have a thermal width, but for heavy quarks
this is expected to be small. Therefore in our study we aim to get the
Green's function in the limit $\epsilon \rightarrow 0$.  In the
numerical analysis we used $\epsilon_c=0.03m_c,~0.01m_c$ and $0.005m_c$
for charmonium, and $\epsilon_b=0.009m_b,~0.003m_b$ and $0.0015m_b$
for bottomonium. For the continuum part of the spectral function all
three values of the width $\epsilon$ give the same result. In the low
energy part the shape of the spectral function agrees quite well for
the smallest two $\epsilon_{c,b}$ values. In what follows, we will
show spectral functions calculated for $\epsilon_c=0.005m_c$ and
$\epsilon_b=0.0015m_b$.  We note that for bound states close to the
threshold even a tiny width could have a significant effect, namely it
could eliminate the bound state peak in the spectral function. But on
the level of correlators this introduces at most a $1\%$ effect.

For the numerical analysis we need to specify the potential in
Eq. (\ref{schroedinger}).  The Cornell parameterization of the
potential turned out to be very successful for the phenomenological
description of the quarkonium spectra, as well as a fit Ansatz for the
lattice data on quark antiquark potential. To include medium effects
at high temperatures, as well as many-body effects at zero temperature
(e.g. threshold for open charm or beauty production, and quarkonia
plus glueball production) we will use the following parameterization
of the potential
\begin{equation}
V(r)= \left\{ \begin{array}{ll}
                   -\frac{\alpha}{r} + \sigma r, & r<r_{med}\\[3mm]   
                   -\frac{\alpha'}{r} e^{-\mu r} + \frac{\sigma'}{\mu} (1 - e^{-\mu r})+V_0, & r > r_{med}
                  \end{array}
          \right.
\label{pot}
\end{equation}
The parameters $\alpha$ and $\sigma$ are fixed by zero temperature
lattice QCD calculations, while other parameters may be temperature dependent. We discuss
the choice of these parameters in the following sections. The
potential used in the numerical analysis is of course smooth. We used
a Fermi Dirac function to interpolate between the two forms in
Eq. (\ref{pot}) at $r=r_{med}$. At finite temperature we also used a more complicated interpolation 
between the short and long distance behavior (see Section \ref{sec:ft} and Appendix C). 
Above deconfinement the singlet free and internal energies of static quark antiquark pair 
have also been used as a potential.

\section{Quarkonium correlators at zero temperature}

In this section we discuss quarkonium correlators obtained from the
spectral functions, which are calculated using a potential model
matched onto the perturbative QCD results at higher energies.

From the spectral function, determined as discussed in
section \ref{sec:nrspf}, we calculate the Euclidean correlators defined
by
\begin{equation}
G(\tau,T)=\int_0^{\infty} d \omega \sigma(\omega,T) K(\omega,\tau,T).
\label{spectral_rep}
\end{equation}
At zero temperature, $K(\omega,\tau,T)=\exp(-\omega \tau)$. We compare
the calculated correlators to recent numerical results calculated from
isotropic \cite{datta04} and anisotropic \cite{jako06} lattice
formulations.

\subsection{Numerical analysis of the spectral functions in the nonrelativistic limit}

Using Eqs. (\ref{green_ps}) and (\ref{green_sc}) we calculate the
pseudoscalar and scalar spectral functions in the nonrelativistic
approximation. To do this, we have to specify the parameters of the
potential $\alpha,~\sigma,~r_{med}, \mu, \alpha',~\sigma'$ and $V_0$, as well as the charm
and bottom quark masses $m_{c,b}$. In this work we are interested in
QCD with only heavy quarks because most of the calculations are performed
in the quenched approximation. In quenched QCD (pure SU(3) gauge
theory) the static quark antiquark potential is well known.  In
particular, lattice calculations of the potential have been
extrapolated to the continuum limit \cite{necco01}.  It turns out that
for distances $r>0.4$ fm an excellent description of the potential
calculated on the lattice can be given by the Cornell parameterization
with $\alpha=\pi/12$ and $\sigma=(1.65-\pi/12)r_0^{-2}$. Here $r_0$ is
the Sommer parameter defined as
\begin{equation}
r^2 \frac{d^2 V}{d r^2}|_{r=r_0}=1.65.
\end{equation}
\begin{table}
\begin{tabular}{|c|c|c|c|c|c|c|}
\hline
\multicolumn{3}{|c|}{Charmonia}        & & \multicolumn{3}{|c|}{Bottomonia}  \\
\multicolumn{3}{|c|}{$m_c=1.19$ GeV}   & & \multicolumn{3}{|c|}{$m_b=4.575$ GeV}  \\
\hline
State        &  Model   & Lattice      &&     State    &  Model       & Lattice  \\
\hline        
$1^1S_0$  & 3030        & 3012(1)      &&   $1^1S_0$   & 9406         & 9400     \\ 
$1^3S_1$  & 3030        & 3084(1)      &&   $1^3S_1$   & 9406         & 9426(4)  \\ 
$1^3P_0$  & 3437        & 3408(9)      &&   $1^3P_1$   & 9736         & 9800(16) \\
$2^1S_0$  & 3675        & 3739(46)     &&   $2^3S_1$   & 9874         & 9938(21) \\
$2^3P_0$  & 3966        & 4008(122)    &&   $2^1P_1$   & 10100        & 10181(64)\\
\hline 
\end{tabular}
\caption{Charmonium and bottomonium masses in MeV calculated in our
  model and in quenched lattice simulations. The values of the charm
  and bottom quark masses in our model are also shown.}
\label{tab:masses_T=0}
\end{table}
As is done in most of the quenched QCD studies, we use the
phenomenological value of the Sommer parameter $r_0=0.5$ fm. The
Cornell parameterization with the above parameters gives a fairly good
description of the lattice data, even at short distances down to $0.1$
fm. Only at distances $r<0.1$ fm the effect of the running coupling
appear to be important \cite{okacz04}. Therefore the Cornell
parameterization is appropriate for describing the quarkonium
spectrum, which is sensitive to the potential in the region $0.1 {\rm
  fm} < r < 1 {\rm fm}$.  The charm and bottom quark masses are chosen
such that the potential model reproduces the quenched lattice data on
charmonium \cite{okamoto02} and bottomonium spectra
\cite{davies94}. Since the NRQCD calculation of Ref. \cite{davies94}
does not give the absolute value of the bottomonium masses, we require
that the mass of $\eta_b$ is equal to $9.4$ GeV.  In table
\ref{tab:masses_T=0}, we show the masses of different quarkonium
states and the values of the quark masses.
\begin{figure}[htb]
\includegraphics[width=9cm]{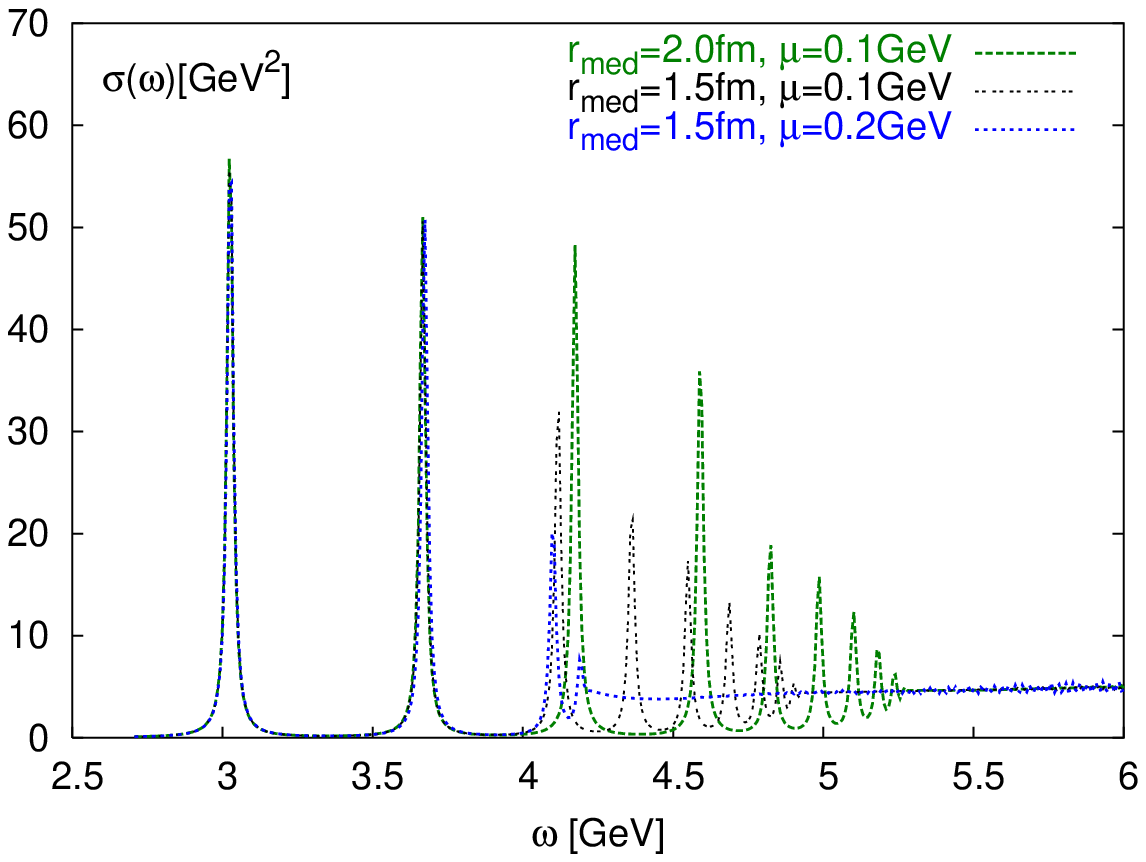}
\includegraphics[width=9cm]{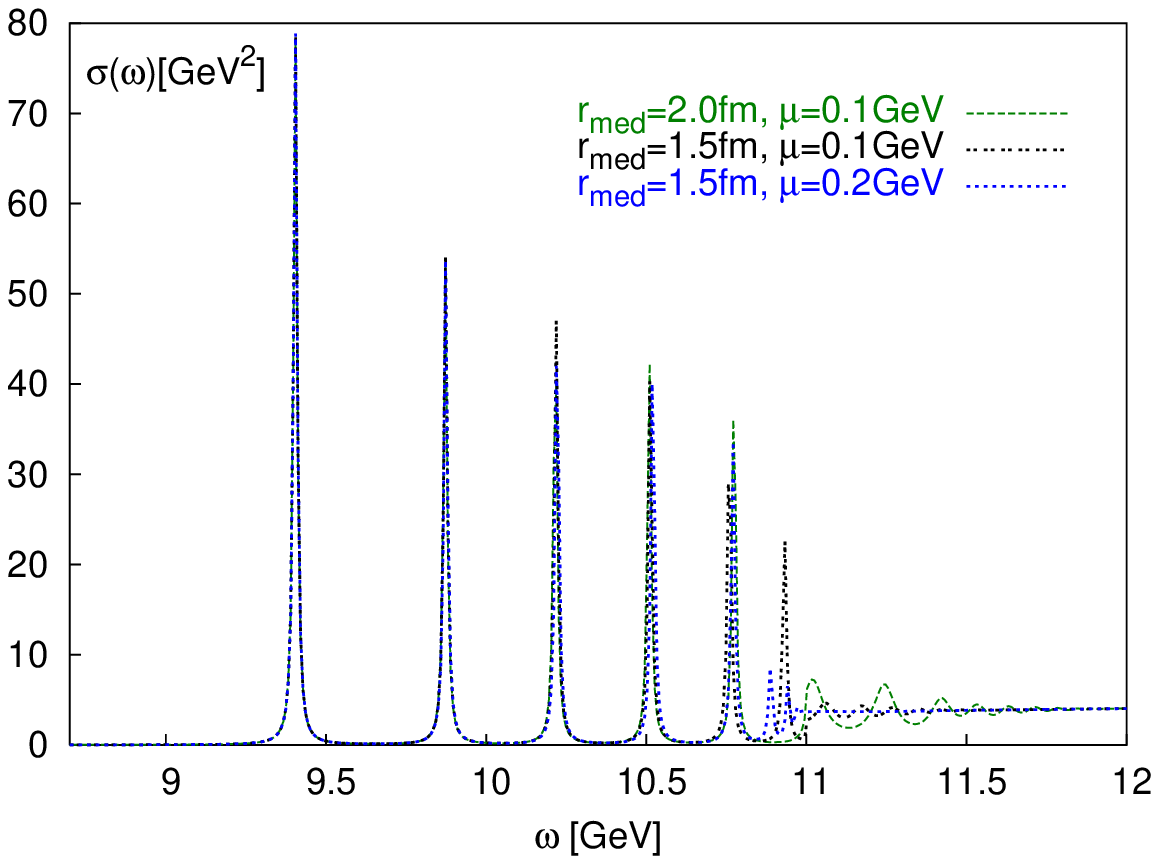}
\caption{The nonrelativistic pseudoscalar spectral functions
  calculated for charmonium (top) and bottomonium (bottom) using the
  screened Cornell potential.}
\label{fig:ps_spf_T=0}
\end{figure}
In Fig. \ref{fig:ps_spf_T=0} we show the quarkonium spectral function
in the pseudoscalar channel for three different sets of the
parameters $\mu$ and $r_{med}$.  The potential (\ref{pot}) corresponding to these
parameter sets is shown in Fig. \ref{fig:pot0}. Here we use $\alpha'=\alpha$ and
$\sigma'=\sigma$.
\begin{figure}[htb]
\includegraphics[width=9cm]{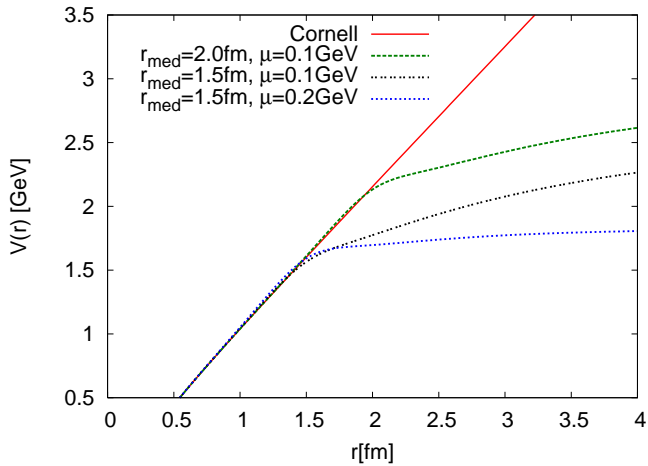}
\caption{The potential at $T=0$ for different parameters. }
\label{fig:pot0}
\end{figure}
In QCD with only heavy quarks, string breaking does not occur until
distances where the potential becomes comparable with twice the heavy
quark mass. However, as the energy $\omega$ is increased the distance
between bound state peaks becomes very small (as can already be seen
in Fig. \ref{fig:ps_spf_T=0}).  Furthermore, at even larger energy, it
is possible to create quarkonium plus glueball states which
subsequently decay into quarkonium states. In this energy regime, it
is impossible to discriminate between individual states and thus the
spectral function will have a continuum. This will happen at energies
of about $\omega \simeq 4-5$ GeV for charmonium and $11$ GeV for
bottomonium.  In this energy region the potential model, strictly
speaking will break down, but the effect of the interaction will still
be important. To mimic the continuum part of the spectral functions,
we will choose $\mu$ and $r_{med}$ such that above energies $4-5$ GeV
for charmonium and $11$ GeV for bottomonium, the corresponding
spectral functions have a continuum. As we will see in the next
section, the correlation function is not very sensitive to the exact
choice of $\mu$ and $r_{med}$ as long as the continuum threshold is
larger than $4$ GeV for charmonium and $11$ GeV for bottomonium, as
shown in Fig.  \ref{fig:ps_spf_T=0}.  This figure illustrates that in the
range studied, since the potential is only modified at distances
larger than $1.5$ fm, the lowest lying states are not affected by the
choice of the parameters $r_{med}$ and $\mu$. In summary, many body
effects in the spectral function can be simulated by the screened
Cornell potential given by Eq. (\ref{pot}) with appropriately chosen
$\mu$ and $r_{med}$. As it is discussed in the next subsection this
somewhat ad-hoc treatment of many body effects has almost no effect
on the correlation functions in Euclidean time. 
The width of the quarkonium states in the spectral functions shown in figure \ref{fig:ps_spf_T=0}  are a numerical artifact due to the 
nonzero value of the parameter $\epsilon_{c,b}$. As mentioned in the previous  section, however, this parameter does not have a visible effect on the correlation function.

\subsection{Direct comparison of the correlation functions with lattice data}

The main focus of this paper is to study the temperature dependence  of the correlation function. 
For this the form of the zero temperature correlator is not crucial. 
To ensure, however, that the comparison of the lattice data and the potential model is meaningful, it is desirable to show 
that the correlation functions calculated in the model agree at least semiquantitatively with the lattice data. 
In this subsection we compare the corelators from the model calculations to the lattice data at zero temperature.
Quarkonium correlators have been
studied in isotropic and anisotropic lattice formulations. The
correlators of the meson currents calculated on the lattice require
renormalization. The corresponding renormalization constants have been
calculated for isotropic lattices only: see discussion in
Refs. \cite{datta04,datta_sewm04,datta_panic05}.  Therefore, for
comparison with our model predictions, we use the data obtained on
isotropic lattices \cite{datta04}. We use the value of the
renormalization constants given in Ref. \cite{datta04}.  In the case
of charmonium, we compare our calculations to the new lattice
calculation on a $48^3 \times 64$ lattice at $\beta=6/g^2=7.192$
\cite{datta_tbp}. In the bottomonium case, we compare our calculations
to the lattice data of Ref. \cite{datta_panic05}.  In these studies
the quark masses, and thus the meson masses, were larger than their
physical value. For this reason, we repeat the analysis from the
previous section using larger quark masses. In Table \ref{tab:mass_is}
we give the resulting quarkonium masses as well as the corresponding
charm and bottom quark mass. In the present analysis, we use $r_0=0.5$
fm as well as the interpolation formula for $r_0$ in the gauge
coupling $\beta$ given in Ref. \cite{necco01} to set the scale. As a
consequence the value of the lattice spacing is smaller; we get
$a=0.017$ fm for the lattice spacing at $\beta=7.192$. Because of
this, the meson masses are larger than those quoted in
Ref. \cite{datta04}, where the value of the string tension
$\sqrt{\sigma}=425$ MeV was used to set the scale.

The relation between the spectral function and the nonrelativistic
Green's function discussed in Section II holds only at leading
order. It will be modified by radiative and relativistic
corrections. The radiative corrections, in particular,  turn out to be quite large
\cite{eichenquig,melnikov,beneke}.
\begin{figure}[htb]
\includegraphics[width=9cm]{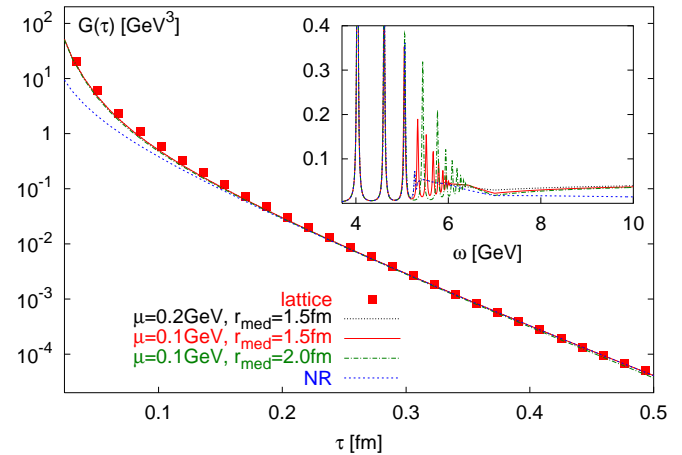}
\caption{The pseudoscalar charmonium correlator calculated in our
  model and compared to lattice data of Ref. \cite{datta04}. In the
  inset, the corresponding spectral functions
  $\sigma(\omega)/\omega^2$ are shown.}
\label{fig:spf_ps_is}
\end{figure}
To take into account these corrections we introduce $K$ factors. These
are determined such that the large $\tau$ behavior of the correlators
calculated in our model matches the lattice data, i.e. we assume
\begin{equation}
\sigma(\omega)=K \frac{2N_c}{\pi} \mbox{Im~} G^{nr}(0,0,E+i \epsilon)
\label{sigma_nr}
\end{equation}
for the $S$-wave quarkonia and similar relation between the derivative
of the Green's function for the $P$-wave quarkonia. The values of $K$
are given in Table \ref{tab:masses_T=0} for each channel. 
For the study of the temperature dependence of the correlator, which is the main objective
of this paper, omitting $K$ would have an effect on the ratio $G/G_{rec}$ which is smaller than  $0.1\%$. 

It is important to be aware that the nonrelativistic approximation
breaks down at large enough energies. On the other hand, for large
energy, the perturbative result for the spectral function given in
Eq. (\ref{free_rel_spf}) becomes reliable. Therefore the
nonrelativistic spectral function in Eq. (\ref{sigma_nr}) has been
smoothly matched onto the perturbative relativistic form given by
Eq. (\ref{free_rel_spf}) and $\alpha_s=0.2$. 
In Fig. \ref{fig:spf_ps_is} we show the pseudoscalar charmonium correlators and the corresponding
spectral functions calculated for different values of $r_{med}$ and
$\mu$.  
The model calculation of the correlators show reasonable agreement
with the lattice data,  and the correlator does not depend  on the values of $r_{med}$ and
$\mu$.
Furthermore, our calculations indicate that even if we use the unscreened Cornell potential, this causes 
a change in the correlation function of less than $0.5\%$. Therefore the ad-hoc treatment of many body effects
discussed in the previous subsection appears to have almost no effect on the Euclidean correlators. 

We have used several procedures to match smoothly the nonrelativistic
function to the relativistic behavior at large energies, and have found
that the differences in the Euclidean time correlator introduced 
by different procedures are less than $0.5\%$.

To obtain the correct $\tau$ dependence of the correlators at short distances, the relativistic form of the spectral function must be used.
The nonrelativistic continuum leads to a correlator which is more than an order of magnitude smaller than the lattice data at short distances,
see Fig.  \ref{fig:spf_ps_is}.
In Appendix B we discuss the additional analysis of the behavior of the 
correlation function at $\tau< 0.1$fm, showing that it is clearly dominated by the relativistic continuum contribution of the spectral functions.
This contradicts the statements made recently in the literature
\cite{wong06,alberico06}, that the lattice correlator do not carry
information about the continuum.

\begin{table}
\begin{tabular}{|c|c|c|c|c|c|c|c|c|}
\hline \multicolumn{4}{|c|}{Charmonia} & &
\multicolumn{4}{|c|}{Bottomonia} \\ \multicolumn{4}{|c|}{$m_c=1.7$
  GeV} & & \multicolumn{4}{|c|}{$m_b=5.8$ GeV} \\ \hline State & Model
& Lattice & $K$ && State & Model & Lattice & $K$ \\ \hline $1^1S_0$ &
3936 & 4023(52) & 2.0 && $1^1S_0$ & 11790 & 11820(81) & 1.8
\\ $1^3S_1$ & 3936 & 4129(105) & 0.8 && $1^3S_1$ & 11790 & 11886(81) &
1.38 \\ $1^3P_0$ & 4319 & 4543(207) & 2.0 && $1^3P_0$ & 12120 &
12295(324) & 2.7 \\ \hline
\end{tabular}
\label{tab:mass_iso}
\caption{Charmonium and bottomonium masses in MeV from isotropic
  lattice calculations compared with the masses calculated in our
  model. Also shown are the $K$-factors for the nonrelativistic
  spectral functions.}
\label{tab:mass_is}
\end{table}
The analysis discussed above for the pseudoscalar charmonium correlator we 
repeated in the vector and scalar channels, as well as for the case of bottomonium.
Also in these channels a reasonable agreement between the lattice data and the
model calculations has been found. 

SInce the Euclidean correlators fall off rapidly with inreasing $\tau$ it is difficult
to judge the agreement between the lattice data and the model calculations only looking
at Fig. \ref{fig:spf_ps_is}.  It is better to study the ratio of the correlators calculated on the
lattice to the correlators calculated in our model. This comparison reveals some discrepancies
between the lattice data and the model. 
These discrepancies are due to lattice artifacts (at short Euclidean time separation), 
the limited validity of the nonrelativistic approach, and in some cases due to the lack  
of fine-tuning of the K-factor, as discussed in Appendix B where we show the
details of this analysis as well as our results in other channels.

\subsection{Zero temperature spectral function as reference}
In the previous subsection we have shown that the nonrelativistic
spectral function scaled with proper factors, which take into account
the relativistic and radiative corrections, and matched to the
perturbative relativistic spectral function can give a fairly good
description of the lattice data obtained at quark masses larger than
the physical values. We have repeated the procedure described in the
previous section for the physical value of the quark masses, namely
$m_c=1.19$ GeV and $m_b=4.575$ GeV. In this analysis, we used the
$K$-factors listed in Table \ref{tab:mass_is} and $r_{med}=1.5$ fm and
$\mu=0.2$ GeV.  The calculated spectral function will serve as a
reference against which the finite temperature results in the next
sections will be compared.

\section{Quarkonium correlators at finite temperature}
\label{sec:ft}

As the temperature is increased, quarkonium spectral functions will
change. This eventually results in a temperature dependence of the
Euclidean correlator $G(\tau,T)$.  However, the temperature dependence
of $G(\tau,T)$ is also caused by the temperature dependence of the
integration kernel in Eq. (\ref{spectral_rep}), which at finite
temperature has the form
\begin{equation}
\displaystyle
K(\omega,\tau,T)=\frac{\cosh \omega (\tau-1/(2T))}{\sinh \left( \omega/(2T) \right)}.
\label{kernel_T}
\end{equation}
To separate out the trivial temperature dependence due to the
integration kernel following Ref. \cite{datta04}, we calculate the
reconstructed correlator
\begin{equation}
G_{rec}(\tau,T)=\int_0^{\infty} d \omega \sigma(\omega,T=0) K(\omega,\tau,T)
\label{grec}
\end{equation}
and study the ratios $G(\tau,T)/G_{rec}(\tau,T)$. If the spectral
function does not change across the deconfinement phase transition
this ratio will be unity and independent of the temperature.  In this
Section we discuss the the pseudoscalar and scalar channels in
detail. In the scalar channel, there is a zero-mode contribution above
the deconfinement temperature, i.e. there is a term proportional to
$\omega \delta(\omega)$ in the spectral function \cite{umeda07}.  This
contribution is not present in the derivative $G'(\tau,T)$ of the
correlator $G(\tau,T)$ with respect to $\tau$. Therefore, in the scalar
channel we will consider the ratios of the derivatives
$G'(\tau,T)/G_{rec}'(\tau,T)$ instead of $G(\tau,T)/G_{rec}(\tau,T)$.

\subsection{Correlators in the free case}
At sufficiently high temperatures all quarkonium states will melt and
the interaction between the heavy quark and antiquark will be
weak. In this limit, quarkonium spectral functions are well
approximated by the leading perturbative (free field) expression
\begin{equation}
\sigma_{free}(\omega,T)=\frac{N_c}{8 \pi^2} \omega^2  \left( a + b \frac{s_0^2}{\omega^2} \right )\tanh{\frac{\omega}{4T}}
\sqrt{1-\frac{s_0^2}{\omega^2}}
\label{sigma_free_ft}
\end{equation}
with $s_0=2 m_{c,b}$.  To obtain an estimate of the temperature
dependence of the correlators, we calculate it using $\sigma_{free}$
as the spectral function and consider the ratio $G/G_{rec}$. This
should provide some upper bound on the size of the temperature
dependence of the correlators. In Fig. \ref{fig:gpgrec_free}, we show
the ratios of the correlators and the ratios of their derivatives in the scalar channel. 
In the numerical analysis we have multiplied $\sigma_{free}$ by the
factor $1+C \alpha_s/\pi$ used in Eq. (\ref{free_rel_spf}) to mimic
the leading perturbative corrections at large $\omega$. This form of
the continuum ensures that $G/G_{rec}$ approaches one for small
$\tau$. As one can see from the figure, the correlators corresponding
to free quark propagation are very different from the zero temperature
correlators, namely they are considerably smaller.  The differences
are considerably larger for bottomonium than for charmonium and
slightly larger in the scalar channel than in the pseudoscalar one.
\begin{figure}[htb]
\includegraphics[width=9cm]{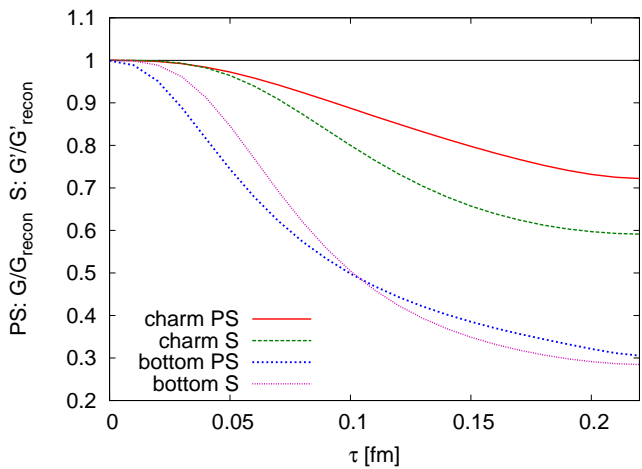}
\caption{The ratio $G/G_{rec}$ for the pseudoscalar and $G'/G'_{rec}$
  for the scalar channels for charmonium and bottomonium. }
\label{fig:gpgrec_free} 
\end{figure}

\subsection{Screening above deconfinement}
\label{sec:screening}
Above the deconfinement temperature, we expect color screening to take
place. The interaction between the heavy quark and the antiquark will
be modified. We assume, however, that the integral equation
(\ref{schroedinger_mom}) for the nonrelativistic Green's function still
holds at finite temperature but with some temperature dependent
potential $V(r,T)$. On the lattice color screening is studied by determinimg 
correlation function of a color singlet static quark
antiquark pair separated by some distance $r$ in Euclidean time, 
evaluated at $\tau=1/T$. This is an expectation value of two Wilson
lines
\begin{equation}
Tr \langle L(0) L^{\dagger}(r) \rangle = \exp(-F_1(r,T)/T)
\label{f1}
\end{equation}
and defines the so-called singlet free energy $F_1(r,T)$
\cite{philipsen04}. Since this object is not gauge invariant, one has
to calculate it in a fixed gauge. Most of the studies of screening of
static quarks use the Coulomb gauge
\cite{okacz02,digal03,okacz04,petrov04,okacz05,petrov06}. Instead of
using the Coulomb gauge, one can insert a spatial transporter between
the static quark and antiquark, i.e. calculate cyclic Wilson loops.
Studies of cyclic Wilson loops at finite temperature have also been
performed and have given results very similar to the Coulomb gauge
calculations \cite{olaf_unpub}.

Studying $F_1(r,T)$ as a function of the separation $r$ three
different regions can be distinguished: the short distance region, the
intermediate distance region and the long distance region. At
sufficiently short distances, $F_1$ is temperature independent and
coincides with the zero temperature potential. In this region $F_1$
does not depend on the choice of the correlation function used to
calculate it (e.g. Coulomb gauge correlator or cyclic Wilson loop). At
some distance $r>r_{med}$, the singlet free energy becomes temperature
dependent and deviates from the zero temperature potential.  The
$r$-dependence of the singlet free energy in this region depends on
the choice of the correlation function (see discussion in
Ref. \cite{me_hard04}).  The value of $r_{med}$ separating the short
and intermediate distance regions depends on the temperature as
$r_{med}(T)=0.43{\rm fm}/(T/T_c)$ \footnote{ Since the scale in this
  paper is set by $r_0=0.5$ fm the value of $r_{med}(T)$ determined in
  Ref. \cite{okacz04} to be $r_{med}(T)=0.48{\rm fm}/(T/T_c)$ has been
  rescaled.}.  Finally at large distances, $rT > (1.0-1.25)$ the
singlet free energy is exponentially screened \cite{okacz04}
\begin{equation}
F_1(r,T)=F_{\infty}(T)-\frac{4}{3} \frac{\alpha_1}{r} \exp(-\sqrt{4 \pi \tilde{\alpha}_1} T r).
\end{equation}
This feature is independent of the choice of the operator. Moreover,
$F_{\infty}(T)$ is universal and can be extracted from the gauge
invariant Polyakov loop correlator $\langle Tr L(0)  Tr L^{\dagger}(r) \rangle$ 
\cite{okacz02}.  This gives the free energy of
infinitely separated static quark antiquark pairs $F_{\infty}(T)$ which is
smaller than twice the self energy of a heavy quark in the medium
$V_{\infty}(T)$. This is because the free energy contains a negative
entropy contribution $-T S_{\infty}(T)$ \cite{okacz02}. In fact this
entropy contribution dominates at high temperatures, making
$F_{\infty}(T)$ negative for $T>3 T_c$ \cite{okacz02}. We would expect
that ${\rm min}(F_{\infty}(T),0)$ is a lower bound on $V_{\infty}(T)$.
Furthermore, for this reason, at temperatures close to $T_c$ the
singlet free energy provides a lower bound on the potential in the
sense that $V(r,T)>F_1(r,T)$.

We assume that the potential $V(r,T)$ shares the general properties of
the singlet free energy discussed above, i.e.  there is a short,
intermediate and long distance region. In the short distance region,
$r<r_{med}(T)$ we assume that $V(r,T)$ is equal to the zero
temperature potential. At large distances, $rT > 1.25$ we assume that
$V(r,T)$ has the form
\begin{equation}
V(r,T)=V_{\infty}(T)-\frac{4}{3} \frac{\alpha_1}{r} \exp(-\sqrt{4 \pi \tilde{\alpha}_1} \mbox{\it T r}),
\end{equation}
where $\alpha_1$ and $\tilde{\alpha}_1$ were determined in
\cite{okacz04}. In the case of a screened Cornell potential
$V_{\infty}=\sigma/\mu$ (c.f. Eq. (\ref{pot})). In QCD with light
dynamical quarks, string breaking occurs at distances of about $1$
fm. Therefore, assuming $\mu \simeq 200$ MeV we estimate that
$V_{\infty}=\sigma/\mu \simeq 1.1$ GeV, which agrees reasonably well
with twice the binding energy of heavy-light meson $2 E_{bind}=2
M_{D,B}-2 m_{c,b}$. Using this analogy and realizing that the role of
the effective screening mass in our case is taken by $1/r_{med}(T)$, we
estimate $V_{\infty}(T)=r_{med}(T) \sigma$. This way we have specified
the large distance behavior of the potential. In the intermediate
distance range $r_{med}(T) < r < 1.25/T$ we use a Fermi-Dirac function
to interpolate between the short and the long distance behavior, such
that the value of the potential and its first derivative agree at
$r=r_{med}$ and at $r=1.25/T$. The potential constructed this way is
shown in Fig. \ref{fig:potTne0} for several different temperatures
together with the lattice data on the singlet free energy. In Appendix C we give details about the construction of this potential. Let us note that the above choice of
$V_{\infty}(T)$ is also motivated by recent model analysis of the singlet free energy
with dimension two gluon condensate \cite{salcedo}. 

\begin{figure}[htb]
\includegraphics[width=9cm]{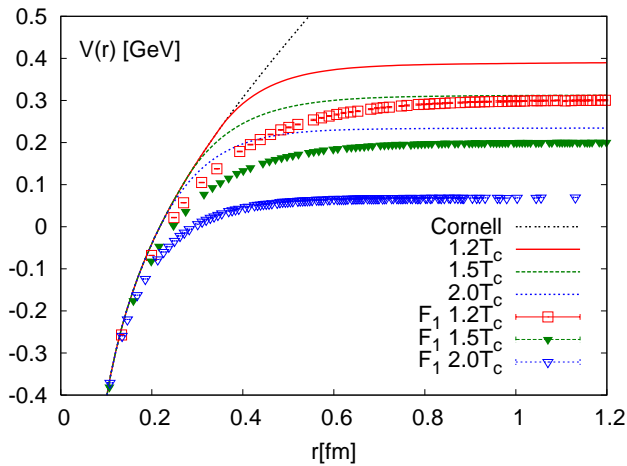}
\caption{The finite temperature potential used in our analysis at
  several values of the temperature together with the lattice data on
  the singlet free energy from
  Refs. \cite{okacz02,okaczlat03,okacz04}}.
\label{fig:potTne0} 
\end{figure}

\subsection{Numerical results with $F_1(r,T)$}

From the above discussion it is clear that the singlet free energy
provides a lower limit for the screened potential. Therefore we have
analyzed charmonium and bottomonium spectral functions using the
singlet free energy as the potential $V(r)$ in
Eq. (\ref{schroedinger}).  The numerical results for the charmonium
spectral functions are shown in Fig. \ref{fig:res_f1_ps} (top)
together with the corresponding correlation functions (inset). As one
can see, all charmonium states are dissolved already at
$T=1.2T_c$. This is in agreement with earlier calculations, which used
$F_1(r,T)$ as a potential \cite{digal01b}. The dramatic changes in the
spectral functions are not reflected in the correlation function,
which shows only about a $4\%$ enhancement. This is due to
the fact that even in absence of bound states the spectral function is
enhanced in the threshold region. The enhancement near the threshold
occurs because the interaction between the quark and antiquark is
still important at this temperatures. An even if the potential is assumed
to be equal to its lower limit the correlation function does not decrease
as expected in the noninteracting case. 
\begin{figure}[htb]
\includegraphics[width=9cm]{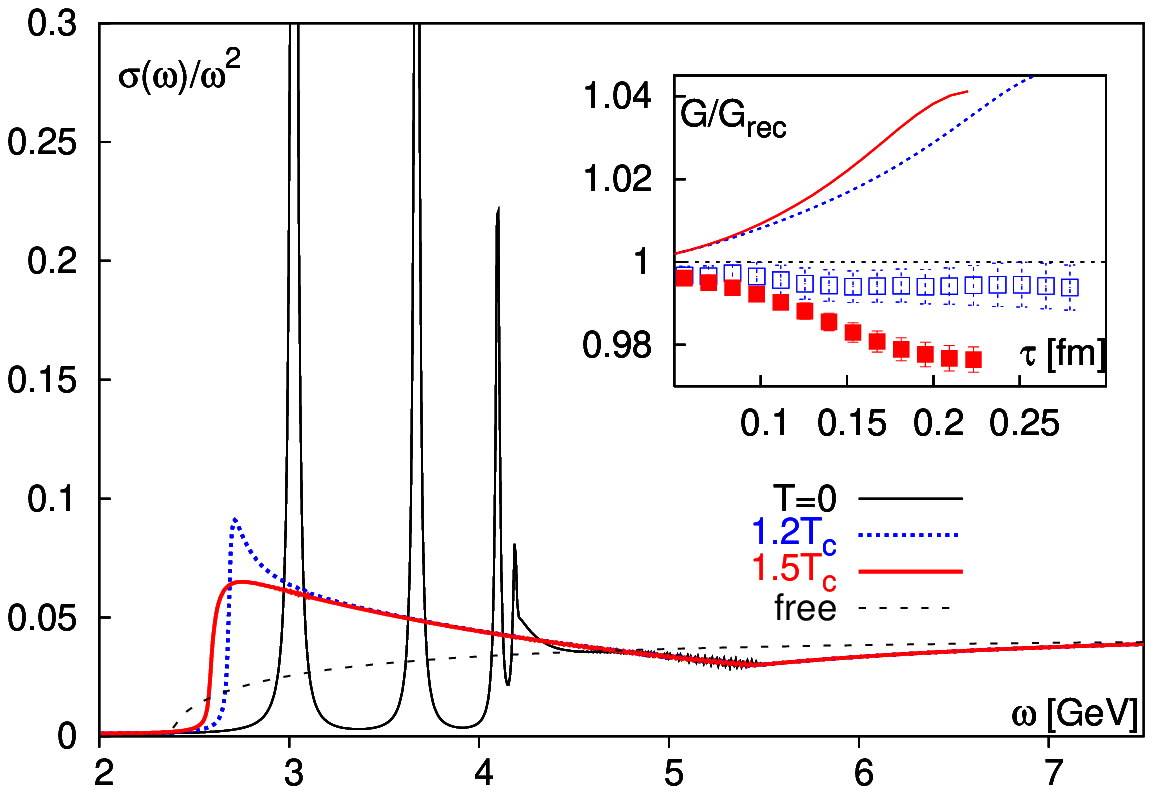}
\includegraphics[width=9cm]{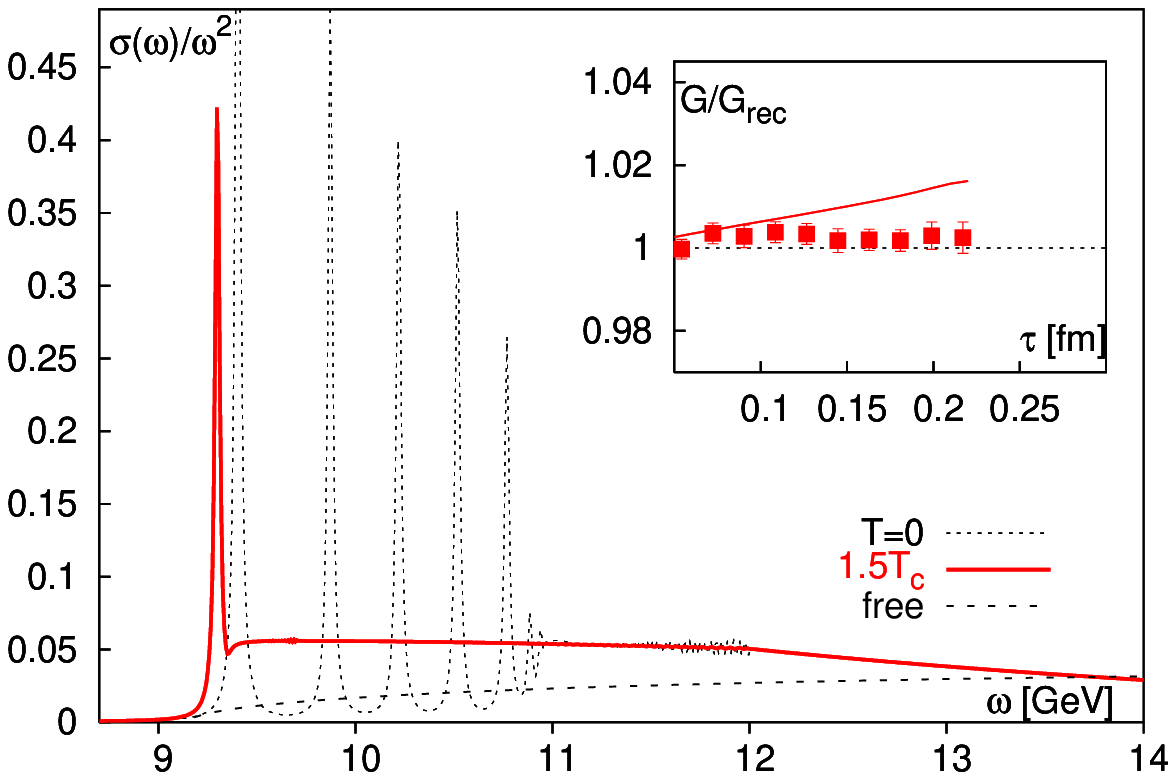} 
\caption{The charmonium (top) and bottomonium (bottom) spectral
  functions at different temperatures calculated using $F_1(r,T)$ as a
  potential. The insets show the corresponding ratio $G/G_{rec}$
  together with the results from anisotropic lattice calculations
  \cite{jako06}.  $G/G_{rec}$ for charmonium lattice data are shown at
  two temperatures $T=1.2T_c$ (open squares) and $1.5T_c$ (filled
  squares), while for bottomonium lattice data at $1.5T_c$ are shown.
}
\label{fig:res_f1_ps}
\end{figure}
In Fig. \ref{fig:res_f1_ps} (bottom) we also show the bottomonium
spectral function and the corresponding correlators. The ground state
bottomonium survives as a resonance up to temperatures as high as
$1.5T_c$. However, due to the shift in the peak position, the
correlation functions at this temperature gets enhanced, resulting in
a small increase in the ratio $G/G_{rec}$.  
The observed enhancement of $G/G_{rec}$ is clearly incompatible with lattice data \cite{datta_panic05,jako06} (see
Fig. \ref{fig:res_f1_ps}).
The analysis of Ref. \cite{blaschke} which uses the free energy as a potential  the melting of charmonium
$1S$ states at $1.2T_c$.  

We have also calculated the spectral function and the correlation
functions in the scalar channel.  The scalar spectral functions show
no resonancelike structure above the deconfinement temperatures
meaning that all P-wave quarkonia are dissolved. The interactions
between the heavy quark and antiquark lead to large enhancement of
the spectral functions, similar to the one observed in the
pseudoscalar channel.  In Fig. \ref{fig:res_f1_sc} we show the ratio
of the derivatives with respect to $\tau$ of $G(\tau,T)$ and
$G_{rec}(\tau)$ for charmonium and bottomonium at $T=1.5 T_c$. Similar
results have been obtained at other temperatures.  Our calculations
agree quite well with lattice results within their rather large
statistical errors. We see again that the large enhancement of the
spectral function near the threshold compensates for the dissolution
of quarkonium states, leaving the correlation function almost
unchanged.
\begin{figure}[htb]
\includegraphics[width=9cm]{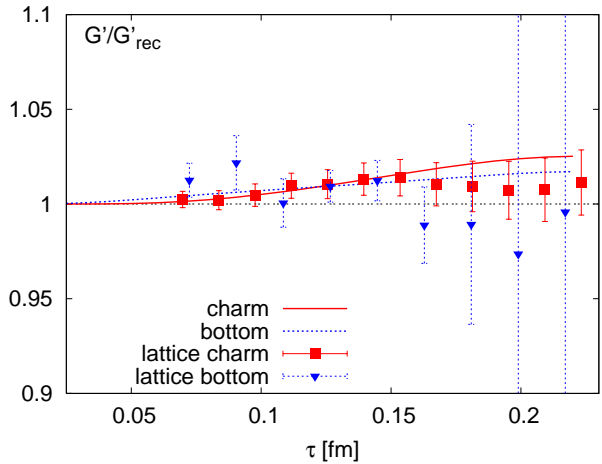}
\caption{ The ratio $G'(\tau)/G_{rec}'(\tau)$ in the scalar channel
  for charmonium and bottomonium at $T=1.5T_c$. Also shown are the
  lattice data for this ratio from Ref. \cite{jako06}.  }
\label{fig:res_f1_sc}
\end{figure}
Thus, contrary to statements made in Refs. \cite{datta04,jako06}, the
dissolution of the $1P$ quarkonium states does not lead to a large
change in the correlation functions. The weak temperature dependence
of $G'(\tau)/G_{rec}'(\tau)$ was first pointed out in
Ref. \cite{umeda07}.  As we will see in section \ref{sec:zeromode}, the
large change in the P-wave correlators observed in
Refs. \cite{datta04,jako06} is due to the zero-mode contribution.

\subsection{Numerical results with the potential $V(r,T)$}
In this subsection we discuss numerical results from our analysis
using the potential $V(r,T)$ discussed in subsection
\ref{sec:screening}. First, we present the pseudoscalar channel. The
charmonium and bottomonium spectral functions at different
temperatures are shown in Fig. \ref{fig:res_v_ps} (top and bottom, 
respectively).  Charmonium spectral functions look similar to the ones
discussed in the previous subsection.  There are no resonance like
structures in the spectral function; only a large threshold
enhancement.  This seems to contradict the conclusions made in
Refs. \cite{umeda02,asakawa04,datta04}, where the analysis of the
spectral function using the MEM indicated that the first peak
survives.  A more detailed analysis of the spectral functions in
Ref. \cite{jako06} resulted in the more modest conclusion that within
numerical accuracy no significant temperature dependence in the
pseudoscalar charmonium spectral functions can be seen.  While
charmonium spectral functions can be reliably reconstructed at zero
temperature using the MEM, this becomes more difficult at finite
temperature due to the fact that the extent of the Euclidean time is
limited to $1/T$ \cite{jako06}.

To understand the situation better, in Fig. \ref{fig:res_v_ps} we show
our results with the lattice charmonium spectral function of
Ref. \cite{jako06} (solid black curve).  The first bump in the
spectral function calculated on the lattice is fairly broad and therefore
its interpretation as the 1S charmonium state is not obvious.
The fact that the bump is centered at $\omega \simeq
3.5$ GeV instead of the expected $\omega \simeq 3$ GeV is a systematic
effect. It has been observed that also at zero temperature, the
position of the first peak is shifted toward larger $\omega$ when
the extent of the Euclidean time used in the analysis is limited to
small values of about $\tau_{max}=0.25-0.3$ fm \cite{jako06}. 
Lattice calculations show, however, that the area under the bump
does not change within the statistical errors  above the deconfinement temperature.
More precisely, the spectral function integrated from 2.7GeV to 4.5GeV does not
change between $T=0$ and $1.5T_c$. 
We also calculated in our model the integrated spectral function in this interval and have found
a change of about $5\%$. 
Note
that the spectral function calculated on the lattice has large
statistical errors.  Thus, it is likely that given the statistical
accuracy of existing lattice data on Euclidean correlators the MEM cannot
distinguish between a threshold enhancement and a true resonance like
structure in the spectral function at finite temperature.

In the case of the bottomonium we see only the ground state above the
deconfinement temperatures, all other states are dissolved. In
comparison with calculations discussed in the previous subsection, we
see that the medium modifications of the first peak are smaller. This
is because the potential $V(r,T)$ is deeper than the singlet free
energy (c.f. Fig. \ref{fig:potTne0}).
 
In the insets in Fig. \ref{fig:res_v_ps} we also display the ratio
$G/G_{rec}$ which shows a much better agreement with the lattice data
compared to the calculations with $F_1$ discussed above.  The
deviations between the lattice data and the results of our
calculations in the charmonium case for $G/G_{rec}$ are less than
$2\%$ for temperatures $T \le 1.5T_c$. As the temperature is further
increased lattice calculations show that $G/G_{rec}$ decreases
monotonically \cite{datta04,jako06}. At $T=3T_c$ its value is about
$0.9$. However, we do not see such large decrease in our calculations.
We expect
that this discrepancy is due to the breakdown of the nonrelativistic
approximation. As the temperature is increased the charmonia bound
states are melted and therefore the typical velocity of heavy quarks
becomes of the order of the thermal velocity $\sqrt{T/m_c}$. At
$T=2T_c$ it becomes $v \simeq 0.7$. Thus the nonrelativistic
approximation breaks down and the simple Ansatz for the
nonrelativistic spectral function given by Eq.(\ref{sigma_nr}) is no
longer valid. As the temperature is increased beyond this point we
expect that the spectral function should slowly approach the free
continuum form (\ref{sigma_free_ft}) and thus the ratio $G/G_{rec}$
should decrease (recall Fig. \ref{fig:gpgrec_free}).  We would expect
that for bottomonium the nonrelativistic approximation should be
valid at higher temperatures and therefore the calculated ratio
$G/G_{rec}$ should agree better with the corresponding lattice
data. Results in Fig. \ref{fig:res_v_ps} show that this is indeed the
case.
\begin{figure}[htb]
\includegraphics[width=9cm]{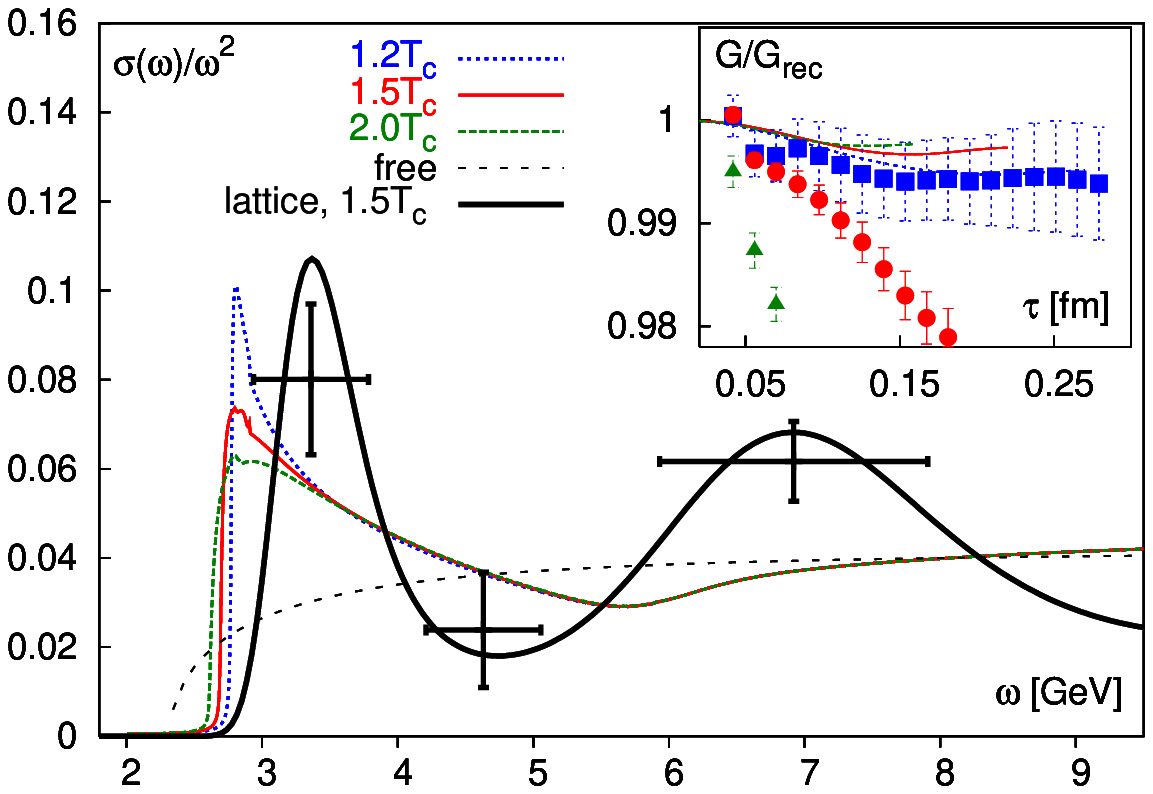}
\includegraphics[width=9cm]{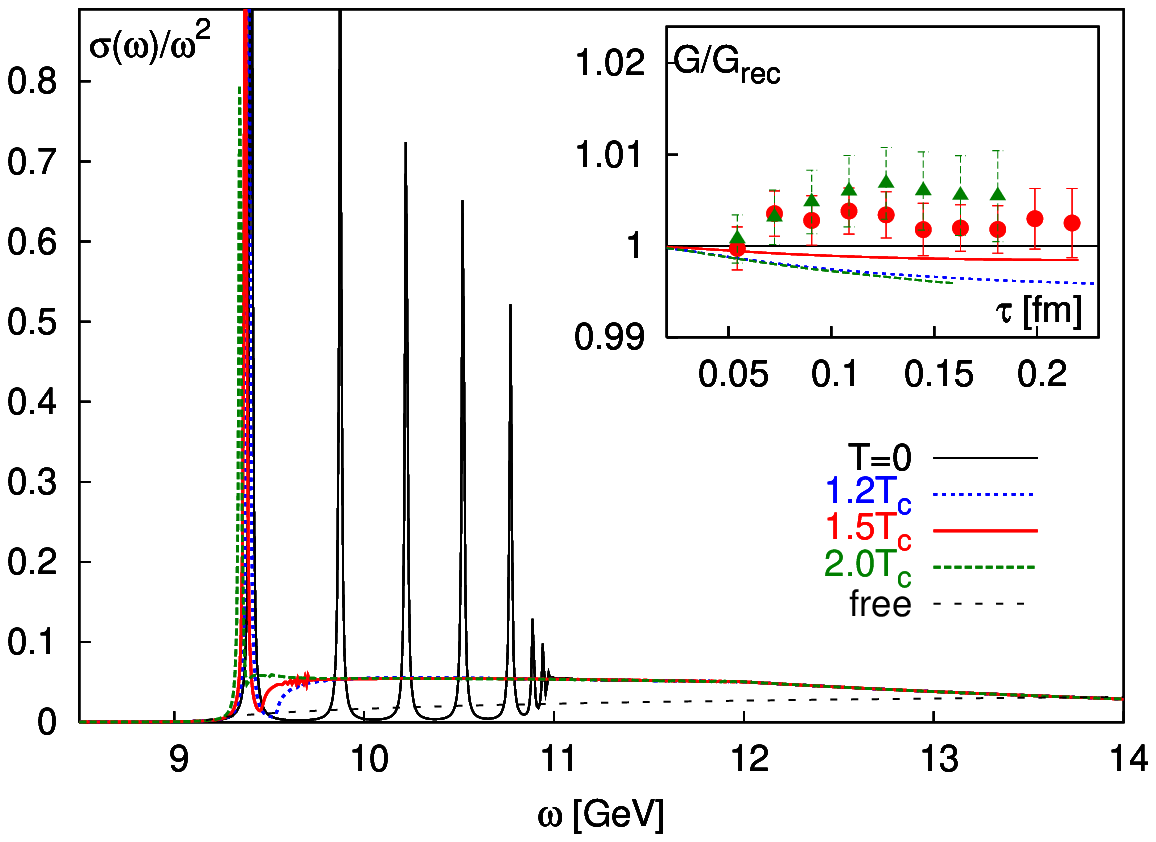} 
\caption{The charmonium (top) and bottomonium (bottom) spectral
  functions at different temperatures calculated using $V_1(r,T)$ as
  the potential. For charmonium we also show the spectral functions
  from lattice QCD obtained from the MEM at $1.5T_c$. The error-bars
  on the lattice spectral function correspond to the statistical error
  of the spectral function integrated in the $\omega$-interval
  corresponding to the horizontal error-bars.  The insets show the
  corresponding ratio $G/G_{rec}$ together with the results from
  anisotropic lattice calculations \cite{jako06}.  For charmonium,
  lattice calculations of $G/G_{rec}$ are shown for $T=1.2T_c$
  (squares), $1.5T_c$ (circles), and $2.0T_c$ (triangles). For
  bottomonium lattice data are shown for $T=1.5T_c$ (circles) and
  $1.8T_c$ (triangles).  }
\label{fig:res_v_ps}
\end{figure}

\subsection{Other choices of the potential}

The choice of the potential discussed above is somewhat
ambiguous. Given the values of $r_{med}(T)$ and $V_{\infty}(T)$ there
are many ways to interpolate smoothly between the short and long
distance regimes. We tried different interpolations and found that the
spectral functions do not depend on the method used. Next, one can ask
how the results depend on the value of $V_{\infty}(T)$. The free
energy $F_{\infty}(T)$ gives a lower bound on this quantity because of
the negative entropy term, and the internal energy
$U_{\infty}(T)$ provides an upper bound.  Therefore we
use the singlet internal energy calculated in Ref. \cite{okaczlat03}
as a potential. This quantity has often been used as a potential when
discussing quarkonium properties at finite temperature
\cite{wong06,alberico,rapp}. The details of the calculations are
discussed in Appendix D.  We find that at $1.2T_c$ the $1S$ charmonium
and $2S$ bottomonium states are present as resonances in the spectral
functions.  At higher temperature, of about $T=1.4T_c$ the only
resonant structure which is present is the $1S$ bottomonium state.  We
also see that the properties of the ground state are significantly
modified for this choice of the potential. As a consequence, the
temperature dependence of the charmonium correlators does not agree
with the lattice results. We also find significant deviation in the
bottomonium correlators at $2T_c$. Thus the singlet internal energy is
not a reasonable choice for the potential. 

The value of $r_{med}(T)$
in the above analysis was chosen according to the analysis of the
singlet free energy. There are many states which contribute to the
singlet free energy. As such, the onset of a temperature dependence in
the free energy at some distance does not necessarily imply a strong
temperature dependence for the potential. However, $r_{med}(T)$ should
be smaller than the distance where we see exponential
screening. Therefore, we take $r_{med}(T)=1.25/T$ as the upper bound
on $r_{med}(T)$. This gives $r_{med} \simeq 0.7$ fm at $1.2T_c$ and
$r_{med} \simeq 0.5$ fm at $1.5T_c$.  The numerical results obtained
from the potential constructed with these values of $r_{med}$ are
discussed in Appendix D.  We see in particular that with such a choice
for the potential the peaks corresponding to $1S$ charmonium, $1P$
bottomonium, and $2S$ bottomonium resonances are present and unchanged
in the spectral function. The binding energy of these states,
i.e. the distance between the resonance peak and the continuum, are
quite small.  Therefore thermal fluctuations can destroy these
states. At $1.5T_c$ we do not see any resonance like structures,
except the $1S$ bottomonium state.  Therefore we see that for all
choices of the potential which are consistent with the information on
color screening coming from lattice QCD, all quarkonium states, except
the $1S$ bottomonium, are dissociated at temperatures equal or smaller
than $1.5T_c$.

In addition, we have studied quarkonium spectral functions using
potentials which are different from the ones mentioned above. A common
feature of all the potentials is that they have the same short
distance behavior. 
From the discussion in the previous two subsections it is clear that the strong threshold enhancement is 
due to the short distance behavior of the potential.
The long distance part however, is different.
Here we do not use lattice data to constrain the long distance behavior of
the potential.  
The details of the analysis are given in Appendix D.
The quarkonium correlators obtained with these potentials also show a
weak temperature dependence. More precisely, we find that
$G(\tau)/G_{rec}(\tau)$ in the pseudoscalar channel and
$G'(\tau)/G_{rec}'(\tau)$ in the scalar channel are temperature
independent and are close to unity. Thus the temperature dependence
of the correlators does not depend strongly on the details of the
potential.

\section{zero-mode contribution to quarkonium correlators}
\label{sec:zeromode}
So far, when addressing the scalar channel we have discussed only the
derivatives of quarkonium correlators.  The reason for this is the
presence of a zero-mode contribution at finite temperature, i.e.
there is an extra finite temperature contribution at $\omega \simeq 0$
in the quarkonia spectral function
\begin{equation}
\sigma_i(\omega,T)=\sigma_i^{high}(\omega,T)+\chi_i^s(T) \omega \delta(\omega)\, .
\end{equation}
Here $\sigma_i^{high}(\omega)$ is the high energy part of the
quarkonium spectral function discussed in the previous sections,
i.e. the one at $\omega > 2 m_{c,b}$ and $i=sc,vc, ax$ for the scalar
vector and axial-vector channels respectively.  For the vector channel
this has been discussed in Refs. \cite{mocsy06,derek}.  In the
interacting theory the delta function is smeared and has a Lorentzian
form with the width $\eta$ determined by the heavy quark diffusion
constant $D$, i.e.  $\eta=T/M/D$ \cite{derek}. For values of $D$ which
are not too small the contribution of the second term in the above
equation to the Euclidean correlator is given by a constant
$G^{low}_i(T)=T \chi_i^s(T)$.  The susceptibilities $\chi_i^s(T)$ have
been calculated in the free theory in Ref. \cite{aarts05} and
read \footnote{There is a miss-print in Eq. (19) of
  Ref. \cite{aarts05}, $(a_H^{(1)}+a_H^{(2)}) I_1$ should read
  $(a_H^{(1)}+a_H^{(3)}) I_1$; we thank Edwin Laermann and Gert Aarts
  for clarifications of this issue.}
\begin{eqnarray}
& \displaystyle \chi_{sc}^s(T)=\frac{6}{\pi^2}
  \int_0^{\infty} d p p^2 \frac{m_{c,b}^2}{E_p^2} \left ( -\frac{\partial
    n_F}{\partial E_p} \right )\\ & \displaystyle
  \chi_{ax}^s(T)=\frac{6}{\pi^2} \int_0^{\infty} d p p^2 \left(1+
  \frac{2 m_{c,b}^2}{E_p^2} \right) \left ( -\frac{\partial
    n_F}{\partial E_p} \right ),
\label{chis}
\end{eqnarray}
where $E_p=\sqrt{p^2+m_{c,b}^2}$ and $n_F=1/(\exp(E_p/T)+1)$. Adding
the constant contribution to the P-wave correlators calculated in the
previous section, we can now calculate the ratio $G/G_{rec}$ in the
scalar and axial-vector channels. The result of these calculations is
shown in Fig. \ref{fig:gpgrecP} both for charmonium and bottomonium
(top and bottom respectively). Results from isotropic \cite{datta04}
and anisotropic \cite{jako06} lattice calculations are also
displayed. The agreement between our simplified calculations and the
lattice data is quite good. 
Our analysis supports the observation made in Ref. \cite{umeda07} that
the large increase in the scalar and axial-vector correlators is due
to the low energy contribution to the corresponding spectral
functions.
One should keep in mind that calculations
on anisotropic lattices were done at quark masses which are somewhat
heavier than the physical quark masses (c.f. Table
\ref{tab:mass_iso}).  In the bottomonium case we see that there is a
quantitative disagreement between the isotropic and anisotropic
lattice calculations. This is likely due to the fact that the lattices
used in Ref. \cite{jako06} were too coarse for precise determination
of the bottomonium susceptibilities $\chi_{sc,ax}^s(T)$. A constant
contribution to the correlator $G^{low}_i(T)=T \chi_i^s(T)$ exists at
any nonzero temperature both in the confined and the deconfined
phase. In the confined phase the quark number is carried by heavy
charm and beauty baryons (remember that here we consider QCD with only
heavy quarks) and thus the constant contribution is proportional to
$\exp(-3m_{c,b}/T)$. This contribution is very small.  In the
deconfined phase at sufficiently high temperatures, quark number is
carried by quarks and the constant contribution goes like
$\exp(-m_{c,b}/T)$, and is described by Eq. (\ref{chis}), which is
much larger.  The fact that we are able to explain the behavior of the
scalar and axial-vector correlators using the ideal gas expressions
for the corresponding susceptibilities (Eq. (\ref{chis})) implies that
already at temperatures around $1.5T_c$, the deconfined charm and
bottom quarks carry the quark number.  This fact directly supports our
observation that almost all quarkonium states are dissociated at this
temperature.
\begin{figure}[htb]
\includegraphics[width=9cm]{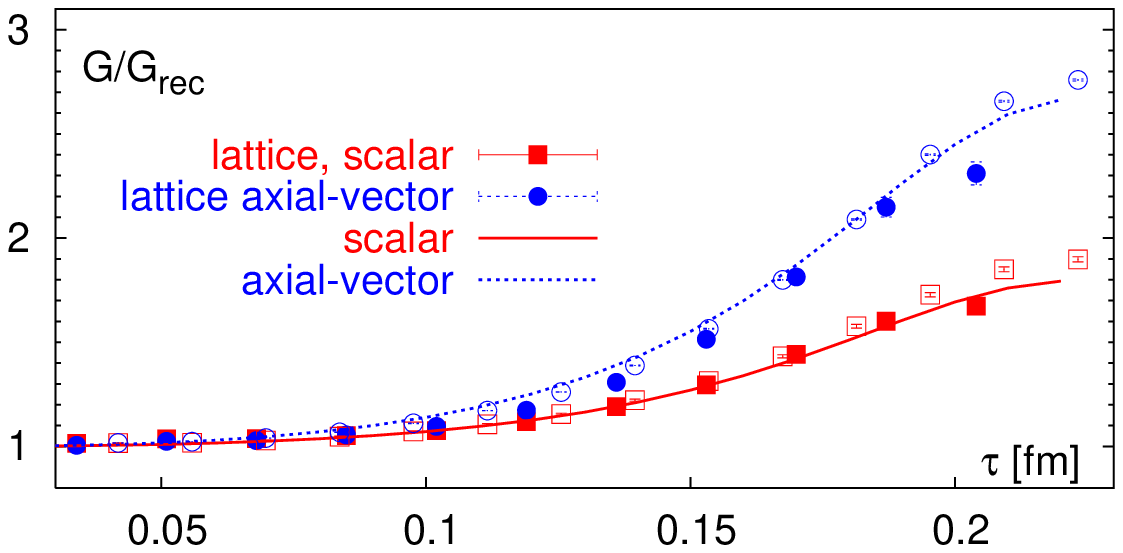}
\includegraphics[width=9cm]{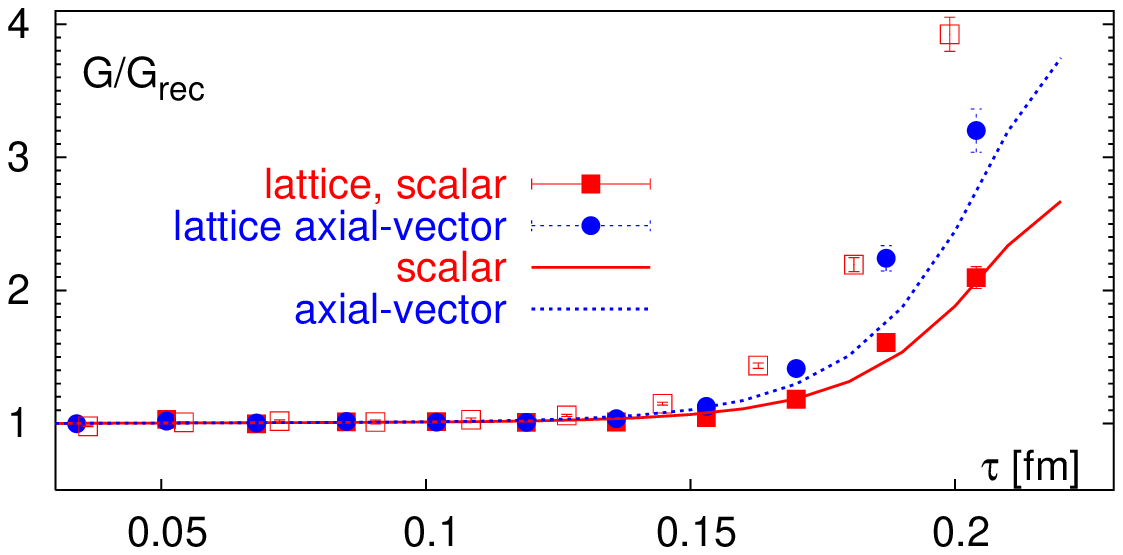}
\caption{The ratio $G/G_{rec}$ in the scalar and axial-vector channel
  at $T=1.5T_c$ for charmonium (top) and bottomonium (bottom).
  Lattice calculation on isotropic lattices
  \cite{datta04,datta_panic05} are shown as filled symbols. Open
  symbols refer to results from anisotropic lattice calculations of
  Ref. \cite{jako06}.}
\label{fig:gpgrecP}
\end{figure}

\section{Discussion and Conclusions}

In this paper we've shown that lattice data on quarkonium correlators
and spectral functions may not necessarily imply survival of different
quarkonium states. We analyzed quarkonium spectral functions by solving
the Schr\"odinger equation for the nonrelativistic Green's function
of a heavy quark antiquark pair. The nonrelativistic Green's
function is expected to describe the spectral function at energies
close to the threshold. The results of these calculations have been
matched to the perturbative form of the spectral functions at high
energies. Although this matching is not unambiguous the Euclidean
correlation functions are not sensitive to the details of the
matching.  We have shown that this very simple approach can give a
reasonable description of the quarkonium correlators calculated on the
lattice.

Let us note that there are several considerations which lead to the
identification of the nonrelativistic Green's function with the
spectral function. One possibility is to construct an effective
theory, the potential NRQCD, where the quark antiquark pair is the
only dynamical field at zero temperature
\cite{Pineda98,Brambilla00,Brambilla05}.  Attempts to generalize this
approach to finite temperatures were recently discussed in
Refs. \cite{laine06,laine07}. Bottomonia spectral functions have been
calculated in this approach using Hard Thermal Loop perturbation
theory resulting in S-wave bottomonia spectral functions which are
similar to ours \cite{laine07}.  Another possibility to relate
Green's functions to the quarkonia spectral function is
to construct an integral equation for the vertex function
$\Gamma(p,q)$ as discussed in section II while systematically taking
into account medium effects.  At sufficiently high temperatures, this
can be done using perturbation theory.  This certainly has to be
investigated in the future.

We calculated quarkonium spectral functions at finite temperature using a class of 
screened potentials based on lattice calculations of the free energy of static quark-antiquark pair. 
Our analysis shows that independent of the details of the choice of the potential,
screening effects lead to dissociation of all quarkonium states with
the exception of the 1S bottomonium state.  
In particular, we find that $1S$ charmonium state dissolves at a temperature
below $1.5T_c$, contrary to the statements made in the literature based on different
potential model analysis \cite{shuryak04,wong04,alberico,wong06,rapp,alberico06} 
as well as on the analysis of lattice spectral functions 
\cite{umeda02,asakawa04,datta04,datta_sewm04,datta_panic05,doi,swan}.
We have analyzed in detail
the temperature dependence of the corresponding Euclidean correlation
functions and have shown that with some reasonable choice of the
potential, the temperature dependence of the quarkonium correlation
functions agrees rather well with the pattern observed in lattice QCD
calculations, i.e. the correlation functions show very little
temperature dependence. Thus lattice data on quarkonium correlators
and spectral functions may not necessarily imply survival of different
quarkonium states. This can be explained by the fact that even in the
absence of resonances, the spectral function is significantly enhanced
in the threshold region compared to the case of free quark
propagation.  This observation may have interesting consequences for
quarkonium phenomenology in heavy ion collisions. It means that a
strong correlation between the heavy quark and the antiquark is
maintained at all temperatures.  The heavy quark antiquark pair is
created in hard processes early in the course of heavy ion collisions
and therefore the separation between the quark and antiquark is small
so that some of them will form correlated pairs. The above observation
implies that the initial correlation between the heavy quark and
antiquark can be maintained, to some extent, through the entire
evolution of the fireball until hadronization. If so, then a fraction of
the correlated quark antiquark pairs emerging from the hard process
can form quarkonium states at hadronization. 

In this paper we discussed the correlation functions of local meson operators. In lattice calculations, 
correlation functions of extended meson operators have also been considered \cite{umeda02,umeda01}.
These could provide further information on the fate of quarkonium states at high temperatures. However,
it is not straightforward to calculate these correlation functions in our approach. Also the comparison
of the lattice data with model calculation is less straightforward, e.g. because of the
renormalization issues.

Lattice calculations of the free energy of a
static quark and antiquark pair show very strong screening effects
\cite{okacz04}. On the other hand, the calculations of the S-wave
quarkonium correlators show very little temperature dependence
\cite{datta04,jako06}. This seemed to be puzzling. In \cite{mocsy05,mocsy06} an attempt to describe quarkonium correlators in potential model has been made. 
In those works the spectral functions consisting of bound state peaks and perturbative continuum has been used, and no agreement with lattice data has been found. 
The same conclusion has been obtained also in \cite{alberico06} when perturbative continuum has been used. More recently, studies which treat bound- and 
scattering-states on an equal footing were presented \cite{rapp,alberico06}. In \cite{alberico06} such treatment improved the agreement with lattice in the pseudoscalar channel. 
In \cite{rapp} no agreement with lattice has been found, which could be due to the fact that the internal energy of a static quark-antiquark pair has been used as the potential. 
To obtain agreement with lattice for the P-wave correlators inclusion of the zero-mode contributions are essential \footnote{Since the present paper has been submitted, 
the role of the zero-mode contribution has been also realized by the authors of \cite{alberico06}, who extended their analysis of the scalar,vector and axial-vector 
correlators \cite{alberico-br}}. The temperature-dependence of the P-wave correlator gives further evidence that most quarkonium states are dissolved above deconfinement.  
In our present work, for the first time, a quantitative understanding of the temperature
dependence of the quarkonium correlators has been obtained within a
potential model with screening. We were able to explain the temperature dependence of the ratio
$G(\tau)/G_{rec}(\tau)$, despite the fact that most states are
dissolved. Good agreement between the lattice data and the potential
model prediction was also reported in Ref. \cite{wong06}. It
is important to understand the differences between our analysis and
that of Ref. \cite{wong06}. First, different quarkonium states were
reported to exist in the quark-gluon plasma. In particular the
dissociation temperature of the $1S$ charmonium state was reported to
be $1.62T_c$ \cite{wong06}. At this temperature, however, the binding
energy is $0.2$ MeV ! A state with such a small binding energy cannot be
detected in the quark-gluon plasma. Second, the authors assumed that
only the ground state contribute to the quarkonium correlation
function. Recent analysis of the quarkonium spectral function in lattice QCD does not
support this assumption \cite{jako06}. Third, in \cite{wong06} the
ratio $G(\tau)/G_{rec}(\tau)$ has been calculated as 
\be
G(\tau,T)/G_{rec}(\tau,T)=\frac{K(M_q(T),\tau,T)}{K(M_q(T=0),\tau,T)}~ , 
\ee 
where $M_q(T)$ is the quarkonium mass and $K(\omega,\tau,T)$ is
the finite temperature kernel defined in Eq. (\ref{kernel_T}). The
above equation, however, does not take into account that the
contribution of the bound state is proportional to the square of the
wave function at the origin (or the corresponding derivative) which
decreases with temperature if screening is present. Thus, in
\cite{wong06} the agreement between the potential model calculations
and lattice results is accidental and due to the oversimplified
approach to the problem.

The analysis presented here can be extended in different
ways. Although the $1S$ bottomonium state seems to survive in the
deconfined phase, this does not necessarily mean that direct
$\Upsilon(1S)$ production in heavy ion collisions is not suppressed.
In the quark-gluon plasma, the $1S$ state will have a thermal width due
to gluon dissociation, which, due to the reduced binding energy, could
be sizable \cite{dima95}. Clearly, the above analysis can be extended
to full QCD. Such an extension is very timely, as new information on
screening of static quarks in the quark gluon plasma becomes available
from large scale lattice QCD simulations at almost physical quark
masses \cite{petrov06}. Finally, it would be interesting to extend the
analysis to finite spatial momenta. We address these questions in \cite{future}.

\section*{Acknowledgments}
We are grateful to 
J. Casalderey-Solana for his contribution at the early stages of this work, especially for providing the
programs for numerical calculations of the nonrelativistic Green's function.
We thank F. Karsch, D. Kharzeev, L. McLerran and
P. Sorensen for careful reading of the manuscript and valuable comments. 
This work has been supported by U.S. Department of Energy under
Contract No. DE-AC02-98CH10886. 

\section*{Appendix A}

In this Appendix we extend the method developed by Strassler and
Peskin in \cite{peskin91} to find, besides the nonrelativistic 
S-wave, also the P-wave Green's function of a central potential.  This
method can be used for any potential which is less singular than
$1/r^2$.

First, we decompose the Green's function into spherical harmonics:
\ber
\label{GofY}
&
G^{nr}(\vec{r},\vec{r'},E+i\epsilon)=\nonumber\\
&
\sum_{l=0}^\infty\sum_{m=-l}^l\frac{g_l(r,r',E+i\epsilon)}{r r'}Y_l^m(\theta,\phi)[Y_l^m(\theta',\phi')]^*
\, ,
\eer
where $\sg$ fulfills the Schroedinger  equation
\ber
\label{d1G}
\left[\frac{d^2}{dr^2} - \frac{l(l+1)}{r^2} + m(E+i\epsilon-V(r))\right]\sg=\nonumber\\
= m \delta(r-r') \, .
\eer
The general solution of this equation can be written as 
\be
\sg=A\,g^l_>(r_>)g^l_<(r_<) \, ,
\label{g}
\ee
where $g_<$ , $g_>$ are the solutions to the homogeneous equation
regular at zero and at infinity, respectively, and $r_<=\min({r,r'})$, $r_>=\max({r,r'})$.  
 The constant $A$ is given by the Wronskian $\mathcal{W}\left(f_1,f_2;r\right)=\left(f_1 f'_2-f'_1 f_2\right)|_r$, 
\be
A=\frac{1}{\mathcal{W}\left(g^l_>,g^l_<;r\right)}\frac{m}{4\pi} \, .
\ee
In order to determine these two solutions we then study the
homogeneous solutions of Eq. (\ref{d1G}) in the vicinity of the
regular singular point $r=0$. For any potential $V(r)$ less singular
than $1/r^2$ there are two solutions, a regular $g^l_0(r)$ and an
irregular $g^l_1(r)$, which behave as \ber g^l_0(r)&=&r^{l+1} + ... \,
,\\ g^l_1(r)&=&\frac{1}{r^l} + ... \, .  \eer Since $g_0(r)$ close to
the origin has the largest degree, we can find a power series solution
of the form \be
\label{g0as}
g^l_0(r)=r^{l+1}\sum_{n=0}^{\infty} a_n r^n \, .
\ee
From this expression we determine the other,  linearly independent solution 
$g_1(r)$ using standard techniques \cite{Arfken}: 
\be
\label{g1as}
g^l_1(r)=g^l_0(r)\int^r dr' \frac{1}{\left(g^l_0(r)\right)^2} \, .
\ee These two solutions are, in general, divergent at $r\rightarrow
\infty$. Since by construction they are linearly independent, we may
write the solutions $g^l_>$ and $g^l_<$ defined in (\ref{g}) as linear
combinations of $g^l_0(r)$ and $g^l_1(r)$. Since $g^l_<$ must be
regular in the origin, we chose \ber g^l_<(r) &=& g^l_0(r) \\ g^l_>(r)
&=& g^l_1(r) + B^l g^l_0(r) \, , \eer where $B$ is defined as \be
B^l=-\lim_{r\rightarrow\infty} \frac{g^l_1(r)}{g^l_0(r)} \, .  \ee In
order to determine B we solve the homogeneous version of equation (\ref{d1G})
numerically. The initial conditions for this are determined using
equations (\ref{g0as}) and ({\ref{g1as}}).  In particular, we
determine Eq. (\ref{g0as}) up to the fifth power and compute
$g^l_0(\delta)$, $g^l_1(\delta)$, and their first derivatives at
$\delta=0.01 {\rm GeV}^{-1}$.

The relationship between $B^l$ and the Green's function (for S-wave),
or its derivative (for P-wave) is obtained from Eq. (\ref{GofY}) and
is \ber
\label{GS}
&
\lim_{r, r'\rightarrow 0} {\rm Im} G^{nr}\left(\vec{r},\vec{r'},E+i\epsilon\right)  \nonumber\\
&
=-\frac{m}{4\pi}  \lim_{r, r' \rightarrow 0} {\rm Im}~ \frac{g^l_<(r)g^l_>(r')}{rr'} \\
&
=-\frac{m}{4\pi}  \lim_{r\rightarrow 0} {\rm Im} \left(\frac{g^0_1(r)}{r}  + B^0\right) = -\frac{m}{4\pi} {\rm Im} B^0 \, \\
&
\label{GP}
\lim_{r, r'\rightarrow 0} {\rm Im}\vec{\nabla}\cdot  \vec{\nabla}' G^{nr}\left(\vec{r},\vec{r'},E+i\epsilon\right) \nonumber\\
&
= -\frac{3 m}{4 \pi} \lim_{r\rightarrow 0} {\rm Im} \left(\frac{g^1_1(r) }{r^2} + 
 \frac{1}{r}\frac{dg^1_1(r) }{dr}+ 3 B^1\right)\,
\eer

Thus the problem of obtaining the Green's function or its derivative
in the origin is now reduced to obtaining the solutions of the
homogeneous equations. Care should be taken though, since the
solutions $g^l_1(r)$ always contain terms that introduce divergences
as $r\rightarrow 0$. The coefficients of these divergences are
determined by the coefficients of the expansions (\ref{g0as}) and
(\ref{g1as}). The first two coefficients do not depend on the long
range part of the potential, while the others depend on the details of
the potential, but can be calculated analytically for a given
potential not more divergent than $1/r^2$. Since the first few
coefficients are real, and the $1/r^n$-like divergent terms do not
contribute to the imaginary part of the Green's function.  The
divergence of the type $g^l_0(r)\ln r$ in case of the S-waves is also
real and thus it will not enter in (\ref{GS}). For P-waves, however,
the coefficient of the $\ln r$ term in (\ref{GP}) has also an
imaginary part proportional to $\epsilon$.  Since the width of the quark
can be with good approximation be taken to be zero, we can take the
limit of $\epsilon \rightarrow 0$, removing this way the divergence that
would otherwise enter the imaginary part. In the numerical analysis,
where we do have a finite width, we solve this problem by explicitly
subtracting the divergence, which we compute from the behavior of
$g^1_1(r)$ near the origin. The exact form of such terms depends on
the exact form of the potential.

\section*{Appendix B}
\label{app-B}
In this Appendix we give further details on the comparison of the
Euclidean correlators calculated in our model to those from isotropic lattices \cite{datta04,datta_panic05,datta_tbp}.
Since the correlation function decays very rapidly with increasing $\tau$ (see Fig. \ref{fig:spf_ps_is}), in Fig. \ref{fig:rat2mod} we show 
instead the ratio of the Euclidean correlation functions calculated on the lattice and in our model for both charmonium
and bottomonium channels. In the charmonium case the model is capable of describing the lattice data within about
$10\%$ accuracy for $\tau>0.2~$fm. 
This is reasonable, as the validity of nonrelativistic approximation is marginal this case.
At smaller separations lattice data deviate from the model prediction by about $30\%$. 
This is expected due to the lattice artifacts discussed in Ref. \cite{karsch03}. For bottomonium we have a similar agreement between the lattice
data and model calculations which extend to smaller $\tau$. The reason for this is that the relativistic continuum part of the spectral function is less important for bottomonium, 
and it becomes visible only at smaller Euclidean time separations.
Note, that the agreement between the lattice data and model calculations in the pseudoscalar channel could be improved by fine-tuning the K-factor.
However, for $\tau<0.1$fm the discrepancy between our model calculations and the lattice data is larger
than for charmonium. This is understood, because the ${\cal O}(ma)$ discretization errors are significantly larger for bottomonium.  

We would like to stress again that at sufficiently small $\tau$  the relativistic continuum part of the 
spectral function is important. To demonstrate
this point better we have subtracted the contribution of the first and
second peaks in the lattice data using the spectral function
reconstructed with the MEM and compared the subtracted correlator to the
one obtained from the free relativistic spectral function.  This is
shown in Fig. \ref{fig:spf_rel_free}. We find a reasonable agreement
between the free relativistic correlators and the lattice data.
\begin{figure}[htb]
\includegraphics[width=9cm]{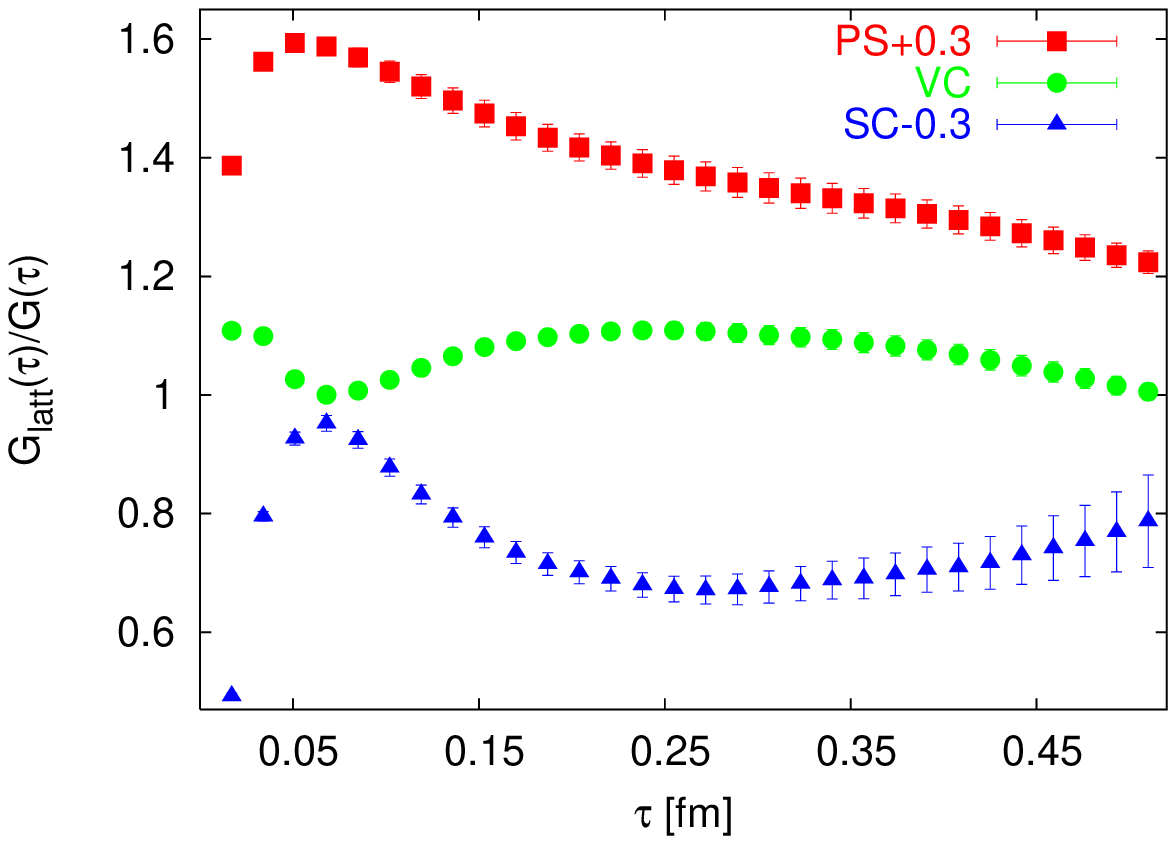}
\includegraphics[width=9cm]{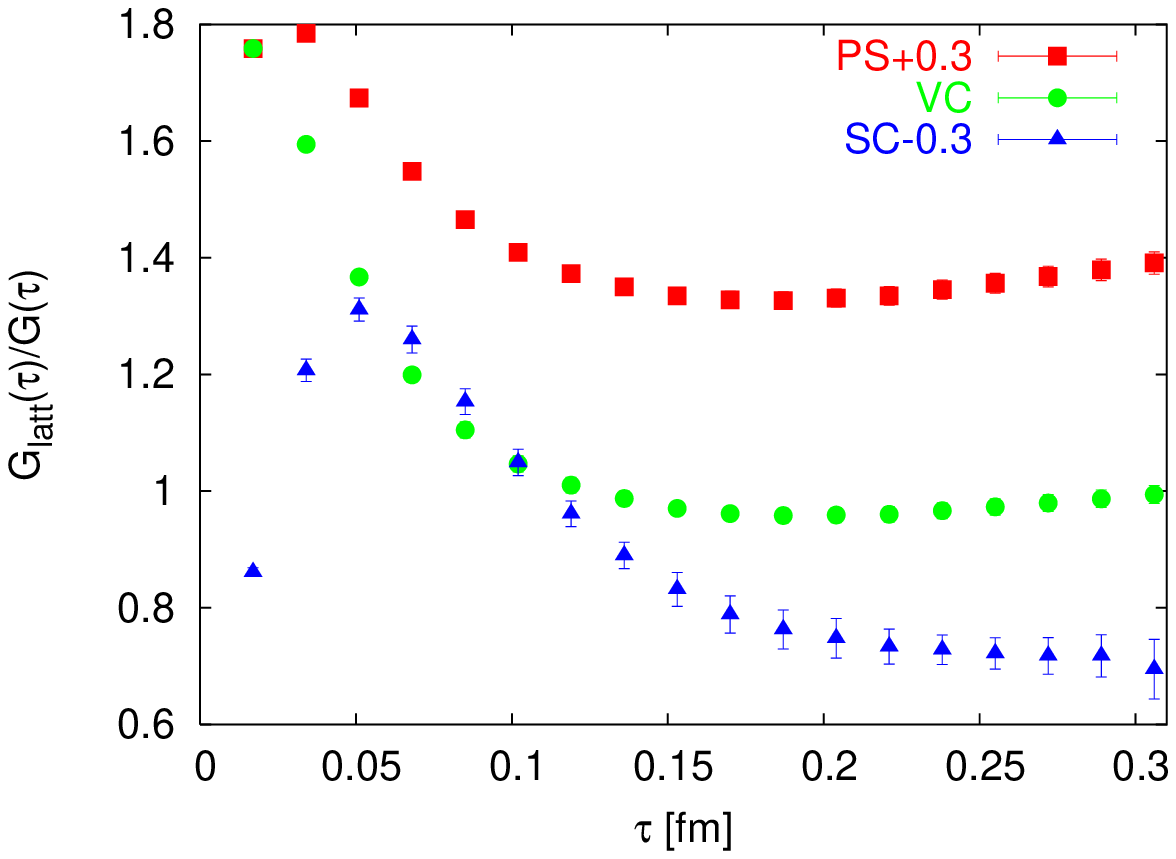}
\caption{The ratio of the correlators calculated on isotropic lattice for charmonium \cite{datta04,datta_tbp} (top)
and bottomonium \cite{datta_panic05} (bottom) to the model calculations in the pseudoscalar, vector and scalar channels.
The data in the scalar and pseudoscalar channel have been shifted by $\mp 0.3$ for better visibility.  
}
\label{fig:rat2mod}
\end{figure}
\begin{figure}[htb]
\includegraphics[width=9cm]{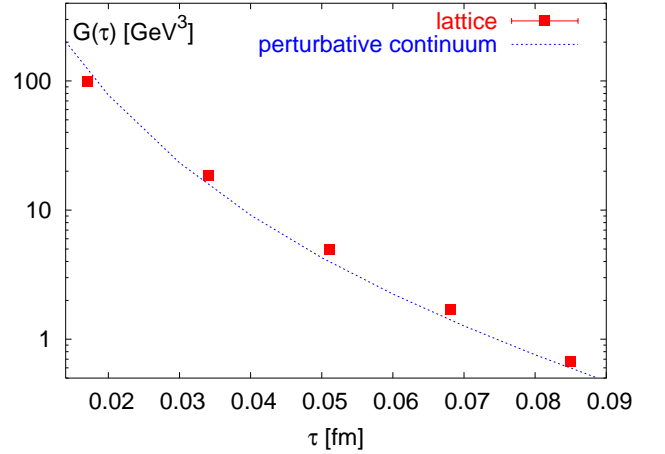}
\caption{The pseudoscalar charmonium correlator calculated on the
  lattice with subtracted resonance contribution compared to the free
  continuum correlator. }
\label{fig:spf_rel_free}
\end{figure}

\section*{Appendix C}
In this appendix we discuss the details of constructing the potential
$V(r,T)$. It has the following behavior in the short, intermediate and
long distance regimes:
 \begin{equation}
 V(r)= \left\{ \begin{array}{l}
V_0(r)=-\frac{\pi}{12}\frac{1}{r} + \sigma r, ~r<r_0=r_{med}\\[3mm]
g(r)= g_1+\frac{g_2}{1+\exp\left(\frac{-r+g_3}{g_4}\right)}+g_5r,~r_{med}<r<r_1 \\[3mm]
V_1(r)= V_{\infty}-\frac{4}{3}\frac{\alpha_1}{r}e^{-\sqrt{4\pi\tilde{\alpha}_1} T r},~r>r_1 
\end{array}\right .
 \end{equation}
with $V_{\infty}=V(r\rightarrow{\infty})=\sigma \cdot
r_{med}(T)$. Furthermore we have \be r_0=\frac{0.43 fm}{T_{red}}
\qquad\mbox{and}\qquad r_1=\frac{1.25}{T},~~T_{red}=T/T_c.
\ee The four parameters $g_1,~g_2,g_3$ and $g_4$ we chosen such that
the value of the function $g(r)$ and of its derivative is equal to the
value of $V_0(r)$ ($V_1(r)$) and the corresponding derivative at
$r=r_{med}$ ($r=r_1$).  Their numerical values are given in Table
\ref{tab:gi}. The value of $g_5$ is zero at $T=1.5T_c,~2.0T_c$, while
$g_5=0.00689804$ at $T=1.2T_c$.  Now we can write the potential for
arbitrary $r$ as \ber
V(r)&=&f_0(r)V_0(r)+(1-f_0(r))f_1(r)g(r)\nonumber\\
&+&(1-f_1(r))V_1(r), \eer where $\quad
f_i(r)=\frac{1}{1+\exp\left(\frac{r-r_i}{\delta}\right)}$ with
$r_i=r_0,r_1$. We choose $\delta=0.001$.
\begin{table}[h]
 \begin{tabular}{||ccccccccc||}
\hline
 $T/T_c$&$r_0$[fm]&$r_1$[fm]&$\alpha_1$&$\tilde{\alpha}_1$&$g_1$&$g_2$&$g_3$&$g_4$\\ \hline 
1.2&0.358333&1.042&1.25&0.5&-2.601&2.982&0.0585&0.0989\\ 
1.5&0.2866&0.833&0.9&0.45&-17.63&17.94&-0.234&0.113\\
2.0&0.215&0.625&0.63&0.35&-1.058&1.292&0.0813&0.0903\\
\hline
\end{tabular}
\caption{The numerical values of parameters which determine the potential $V(r,T)$.}
\label{tab:gi}
\end{table}

\section*{Appendix D}
In this appendix we discuss how our results on quarkonium spectral
functions and correlators depend on the choice of the potential. First,
we use the internal energy $U_1(r,t)$ of static quark antiquark pair
calculated in Ref. \cite{okaczlat03} as the potential. We fit the
lattice data with the following Ansatz \be V(r,T) =
-\frac{\alpha}{r}e^{-\mu r^2} + \sigma r e^{-\mu r^2} +
V_0\left(1-e^{-\mu r^2}\right).
\label{latpot}
\ee 
The fit parameters are given in Table \ref{tab:appC}. In addition
we use a potential given by Eq. (\ref{pot}) for different values of the
parameters $r_{med}$, $\mu$ and $V_0$ and $\alpha'=\alpha$, $\sigma'=\sigma$. 
We have analyzed five different
sets of parameters labeled as II-VI, since the internal energy is
labeled as set I. For set II we set $\mu=\mu(T)=2T$ as predicted by
the lattice QCD and $r_{med}=1.25fm/(T/T_c)$, i.e. we identified
$r_{med}(T)$ with the distance where screening becomes exponential
(see section \ref{sec:screening}). We used the same $r_{med}$ in set
III but $\mu(T)=1/r_{med}(T)$. In parameter sets IV and V we use
$\mu=T$, while in the parameter set VI we used $\mu=2T$.  The
numerical values of $r_{med}$, $\mu$ and $V_0$ for the five different
sets are given in Table \ref{tab:appC}.  
\begin{table}
\begin{tabular}{||ccccc||}
\hline
&$T/T_c$&$\mu$[GeV]&$\sigma$[GeV$^2$]&$V_0$[GeV]\\ \hline
set I &1.13&0.161&0.087&1.094\\
&1.40&0.408&0.000&0.608\\
&1.95&0.891&0.000&0.422\\ \hline \hline
&$T/T_c$&$\mu$[GeV]&$r_0$[fm]&$V_0$[GeV]\\ \hline
set II &1.2&0.708&0.7&0.47\\
&1.5&0.885&0.5&0.2\\ \hline
set III&1.2&0.281&0.7&0.27\\
&1.5&0.394&0.5&0.2\\ \hline
set IV&1.2&0.354&0.557&0.15\\
&1.5&0.4425&0.445&0.10\\ \hline
set V&1.2&0.354&0.40&0.02\\
&1.5&0.4425&0.32&-0.02\\ \hline
set VI&1.2&0.704&0.40&0.03\\
&1.5&0.885&0.32&-0.07\\
\hline
\end{tabular}
\caption{The values of the parameter for different sets of the
  potential used in our analysis.}
\label{tab:appC}
\end{table}
In Fig. \ref{fig:pot1.2_comp}
we show the different choices of the potentials, including the free
energy $F_1(r,T)$ and the potential $V(r,T)$ discussed in section IV,
the internal energy, and the screened Cornell form for the two 
different parameter sets at temperatures of about $T=1.2T_c$.
\begin{figure}[htb]
\includegraphics[width=9cm]{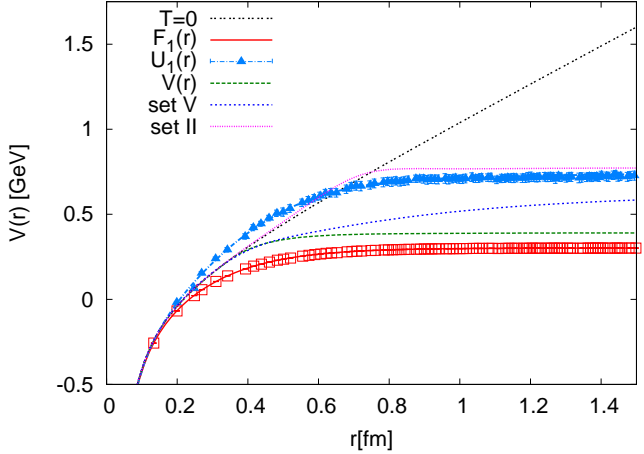}
\caption{Different potential used in the analysis at
  $1.2T_c$. Triangles correspond to lattice data on internal energy
  from Ref. \cite{okaczlat03}, while open squares correspond to
  lattice data on the free energy \cite{okacz02}.  }
\label{fig:pot1.2_comp}
\end{figure}
Below we discuss our numerical results for the internal energy and
different screened Cornell potentials.
\begin{figure}
\includegraphics[width=9cm]{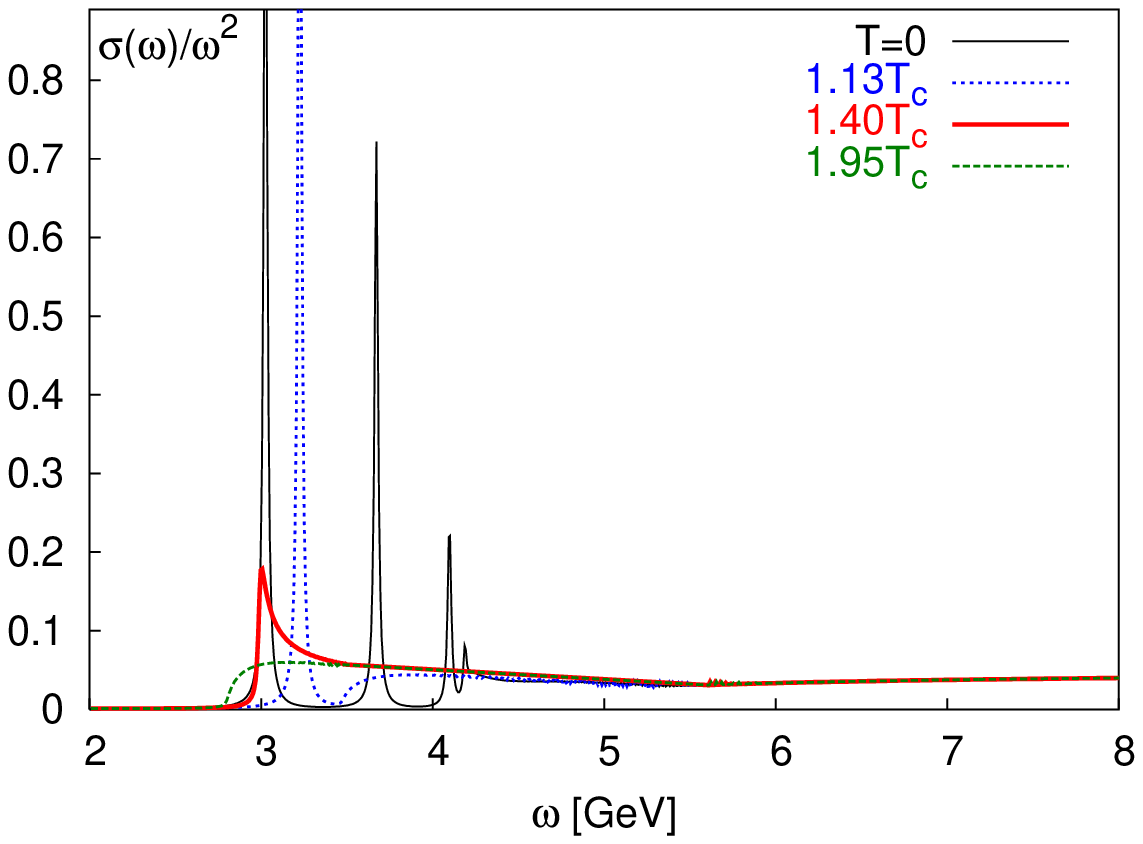}
\includegraphics[width=9cm]{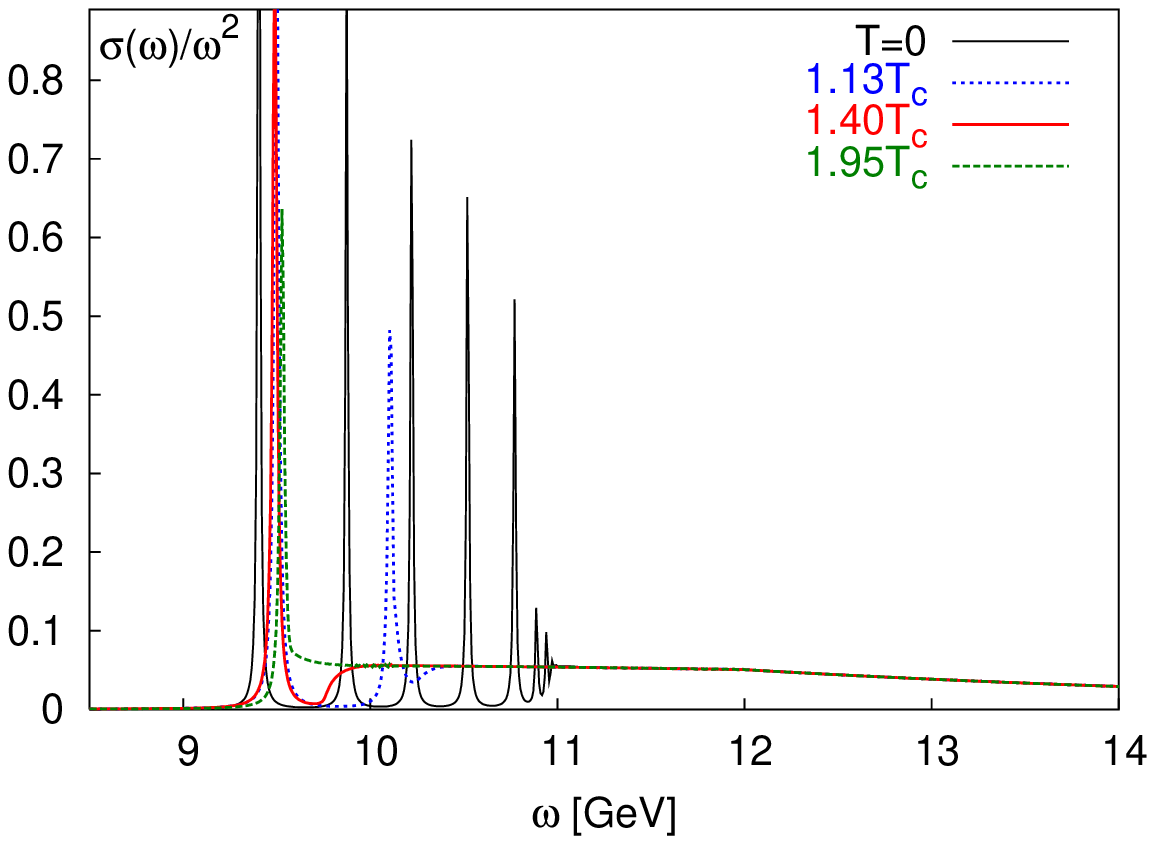}
\caption{Charmonium (top) and bottomonium (bottom) spectral functions calculated
using the internal energy as a potential.}
\label{fig:spf_int}
\end{figure}
The quarkonium spectral functions calculated using the internal energy
as a potential are shown in Fig. \ref{fig:spf_int}.  We see
significant increase in the mass and the amplitude of the $1S$ state
at the lowest temperature for charmonium and all temperatures in the
case of bottomonium. The $1S$ charmonium states dissolves at
temperatures around $1.4T_c$, while the corresponding bottomonium
states exists up to temperatures of about $2.0T_c$. The temperature
dependence of the correlators is shown in Fig. \ref{fig:corr_int}. As
one can see in the Figure, the ratio $G/G_{rec}$ for charmonium is
significantly smaller than unity for all temperatures. In the
bottomonium case $G/G_{rec}$ drops below $0.96$ in contrast to lattice
data. The agreement between the lattice data and potential
model calculations is worst if we use the internal energy as a
potential.
\begin{figure}
\includegraphics[width=9cm]{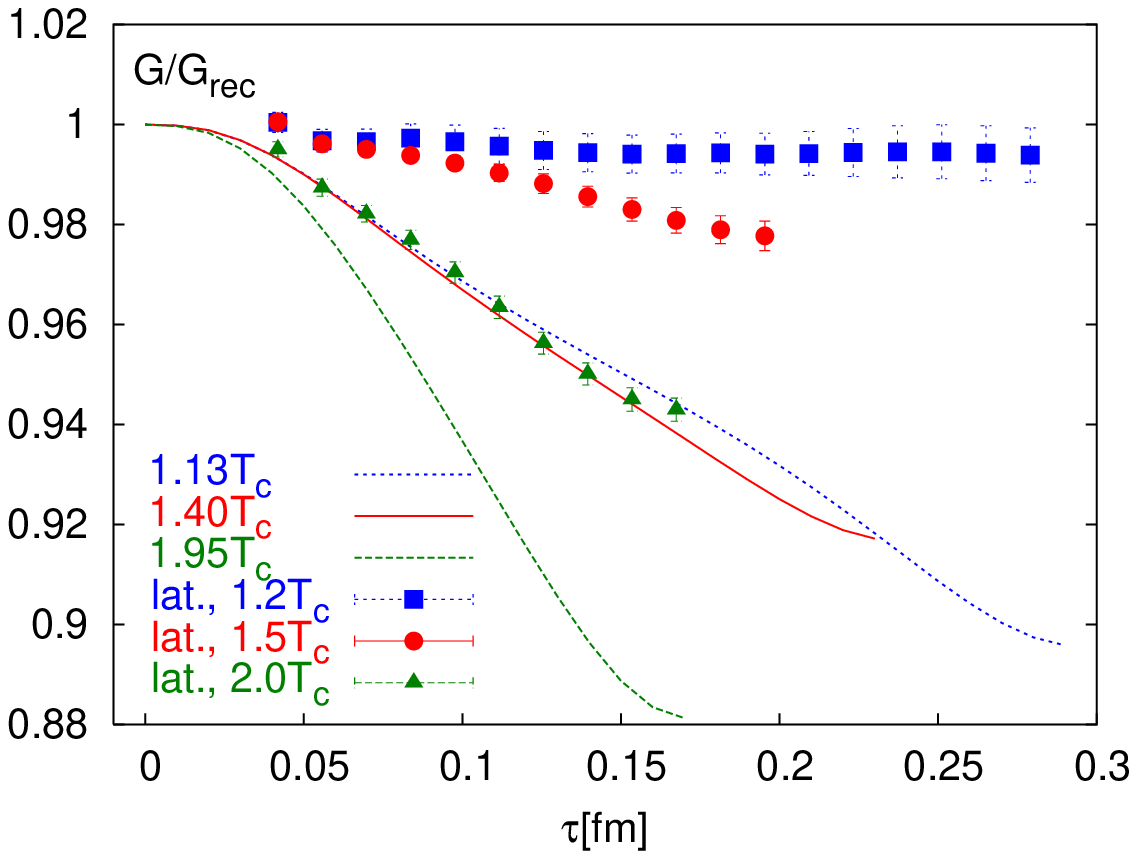}
\includegraphics[width=9cm]{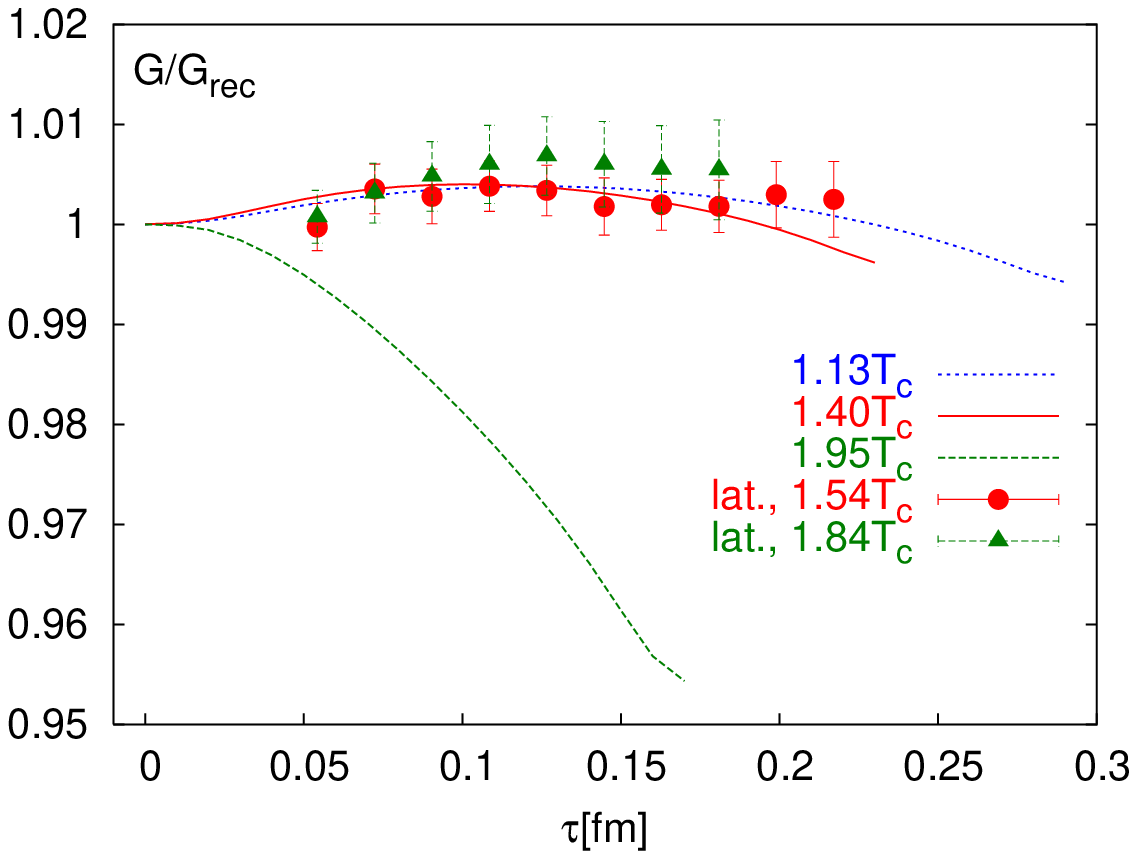}
\caption{The ratio $G/G_{rec}$ for charmonium (top) and bottomonium (bottom) calculated
using the internal energy as a potential.}
\label{fig:corr_int}
\end{figure}
\begin{figure}
\includegraphics[width=9cm]{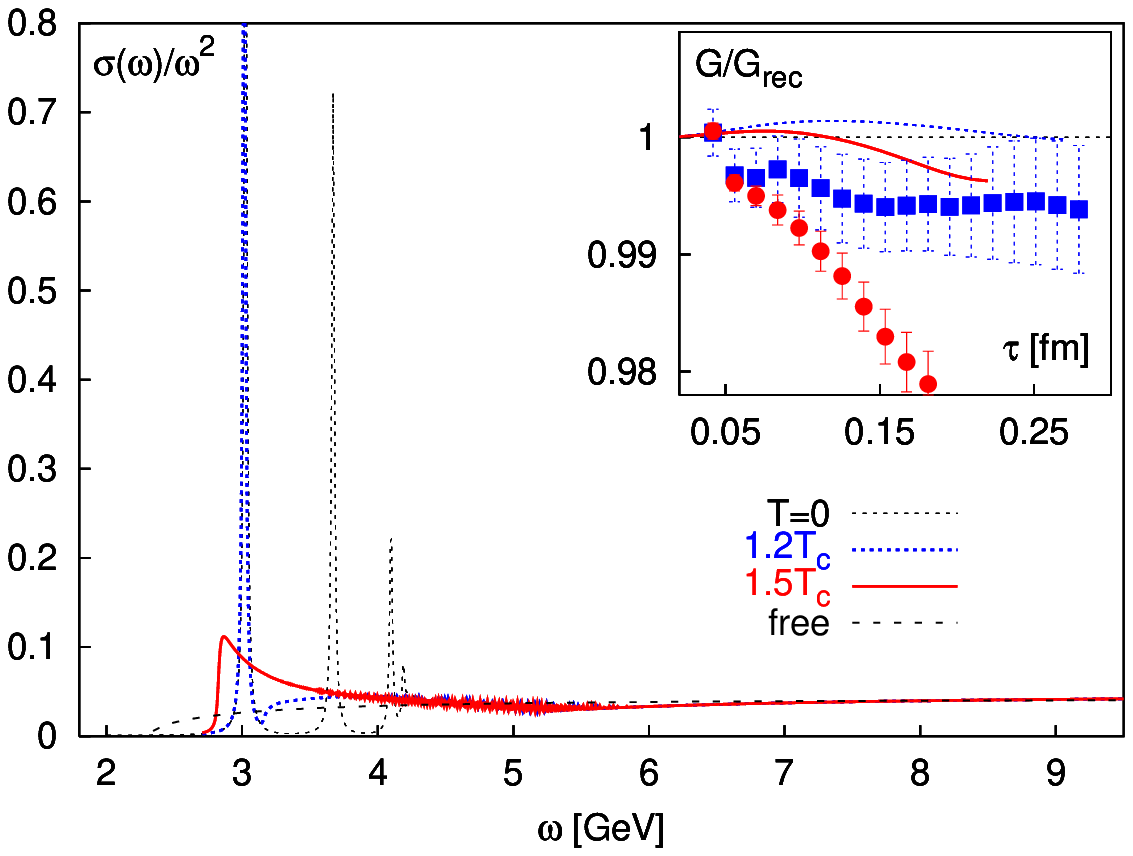}
\includegraphics[width=9cm]{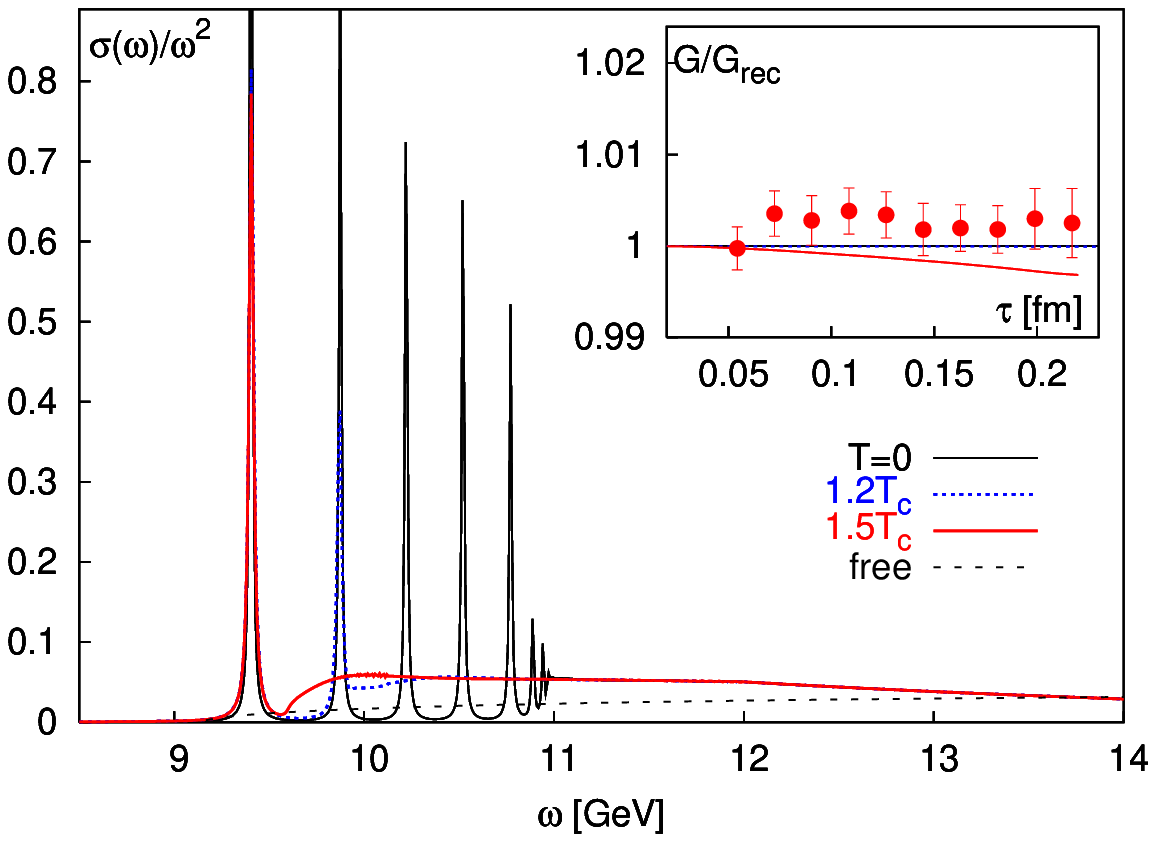}
\includegraphics[width=9cm]{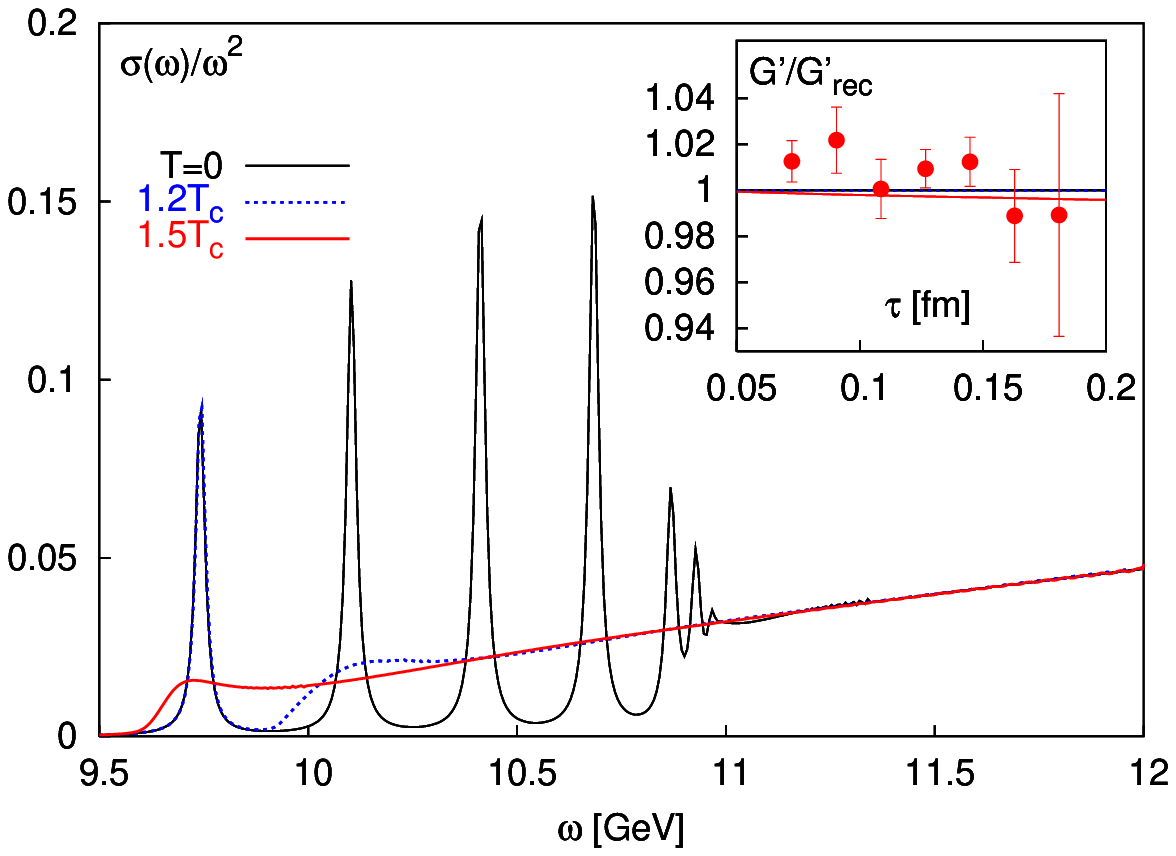}
\caption{The pseudoscalar spectral for charmonium (top) and
  bottomonium (middle) as well as scalar spectral function for the
  bottomonium (bottom) calculated using the potential corresponding to
  set II. In the insets we show $G/G_{rec}$ as well as
  $G'/G_{rec}'$ for the scalar channel. Also shown there are the
  lattice results from Ref. \cite{jako06} at $1.5T_c$. In the case of
  charmonium we also show lattice data at $1.2T_c$ (filled squares). }
\label{fig:spf_set2}
\end{figure}
The quarkonium spectral function for the parameter set II are shown in
Fig. \ref{fig:spf_set2}. As mentioned above the value of $r_{med}(T)$
for this parameter set is about its maximal possible value and
$\mu=2T$ as suggested by lattice QCD.  Therefore quarkonium states can
survive to higher temperatures and have larger binding energies for
this choice of the potential. In this way we can obtain some kind of
upper bound on the dissociation temperatures and binding energies for
different states. We can see from Fig. \ref{fig:spf_set2} that $1S$
charmonium state as well as $2S$ and $1P$ bottomonium states survive
till temperatures as high as $1.2T_c$ for this choice of the
potential. However, the corresponding binding energies are small and
therefore interactions with the medium will result in a sizeable
thermal width and will lead to the dissociation of these states at
this temperatures. At higher temperature, namely $T=1.5T_c$ we no
longer see peaks in the spectral functions corresponding to these
states. Thus charmonium $1S$ state and bottomonium $1P$ and $2S$
states are dissolved at temperatures $1.2T_c < T < 1.5T_c$ for this
potential.  In Fig. \ref{fig:spf_set2} we also show the ratio
$G/G_{rec}$ for the pseudoscalar channel and $G'/G'_{rec}$ for the
scalar channel.  One can see that this ratio is close to one also for
this potential.

We analyzed the spectral functions for other choices of the potential
summarized in Table IV and calculated the ratio $G/G_{rec}$ which are
shown in Fig. \ref{fig:ratcomb} for pseudoscalar charmonium. We see
that for all potentials we have studied this ratio is close to
unity. Similar results have been obtained for pseudoscalar
bottomonium. In the scalar channel we have found the ratio
$G'/G_{rec}'$ is also close to unity.
\begin{figure}
\includegraphics[width=9cm]{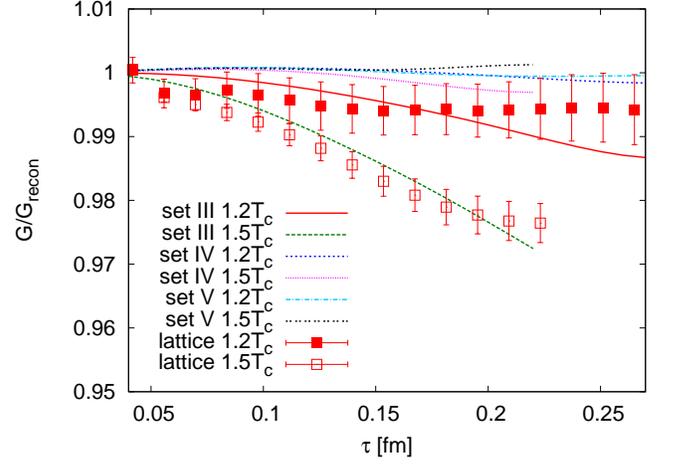}
\caption{The ratio $G/G_{rec}$ for pseudoscalar charmonium for
  different choices of the potential.}
\label{fig:ratcomb}
\end{figure}


\end{document}